\documentclass{article}
\makeatletter
\def\ps@pprintTitle{%
 \let\@oddhead\@empty
 \let\@evenhead\@empty
 \def\@oddfoot{}%
 \let\@evenfoot\@oddfoot}
\makeatother
\usepackage{authblk}
\usepackage[english]{babel}
\usepackage{amsmath,amssymb,graphicx,theorem}

\date{}
\author[1]{
  Phil Weir
}
\author[2]{
   Roland Ellerweg
}
\author[3]{
   Stephen Payne
}
\author[2]{
   Dominic Reuter
}
\author[4]{
   Tuomas Alhonnoro
}
\author[5]{
   Philip Voglreiter
}
\author[1]{
   Panchatcharam Mariappan
}
\author[4]{
   Mika Pollari
}
\author[3]{
   Chang Sub Park
}
\author[6]{
   Peter Voigt
}
\author[7]{
   Tim van Oostenbrugge
}
\author[8]{
   Sebastian Fischer
}
\author[9]{
   Peter Kalmar
}
\author[7]{
   Jurgen Futterer
}
\author[8]{
   Philipp Stiegler
}
\author[9]{
   Stephan Zangos
}
\author[1]{
   Ronan Flanagan
}
\author[6]{
   Michael Moche
}
\author[2]{
   Marina Kolesnik
}
\affil[1]{NUMA Engineering Services Ltd.,}
\affil[2]{Fraunhofer Institute for Applied Information Technology,}
\affil[3]{University of Oxford,}
\affil[4]{Aalto University,}
\affil[5]{Technical University of Graz,}
\affil[6]{Leipzig University,}
\affil[7]{Radboud University Clinic Nijmegen,}
\affil[8]{University Hospital Frankfurt,}
\affil[9]{Medical University of Graz}

\catcode`\<=\active \def<{
\fontencoding{T1}\selectfont\symbol{60}\fontencoding{\encodingdefault}}
\catcode`\>=\active \def>{
\fontencoding{T1}\selectfont\symbol{62}\fontencoding{\encodingdefault}}
\newtheorem{example}{Example}
\newcommand{\mathe}{\mathrm{e}}
\newcommand{\tmem}[1]{{\em #1\/}}
\newcommand{\tmmathbf}[1]{\ensuremath{\boldsymbol{#1}}}
\newcommand{\tmop}[1]{\ensuremath{\operatorname{#1}}}
\newcommand{\tmtextit}[1]{{\itshape{#1}}}
\newenvironment{itemizeminus}{\begin{itemize} }{\end{itemize}}

\begin{document}

\title{Go-Smart: Open-Ended, Web-Based Modelling of Minimally Invasive Cancer
Treatments via a Clinical Domain Approach}

\maketitle

\begin{abstract}
  Clinicians benefit from online treatment planning systems, through off-site
  accessibility, data sharing and professional interaction. As well as
  enhancing clinical value, incorporation of simulation tools affords
  innovative avenues for open-ended, multi-disciplinary research
  collaboration. An extensible system for clinicians, technicians,
  manufacturers and researchers to build on a simulation framework is
  presented. This is achieved using a domain model that relates entities from
  theoretical, engineering and clinical domains, allowing algorithmic
  generation of simulation configuration for several open source solvers.
  
  The platform is applied to Minimally Invasive Cancer Treatments (MICTs),
  allowing interventional radiologists to upload patient data, segment patient
  images and validate simulated treatments of radiofrequency ablation,
  cryoablation, microwave ablation and irreversible electroporation. A
  traditional radiology software layout is provided in-browser for clinical
  use, with simple, guided simulation, primarily for training and research.
  Developers and manufacturers access a web-based system to manage their own
  simulation components (equipment, numerical models and clinical protocols)
  and related parameters.
  
  This system is tested by interventional radiologists at four centres, using
  pseudonymized patient data, as part of the Go-Smart Project
  (\href{http://gosmart-project.eu}{http://gosmart-project.eu}). The
  simulation technology is released as a set of open source components
  (\href{http://gosmart-project.eu}{http://github.com/go-smart}).
\end{abstract}

\begin{figure}[h]
  \resizebox{340px}{!}{\includegraphics{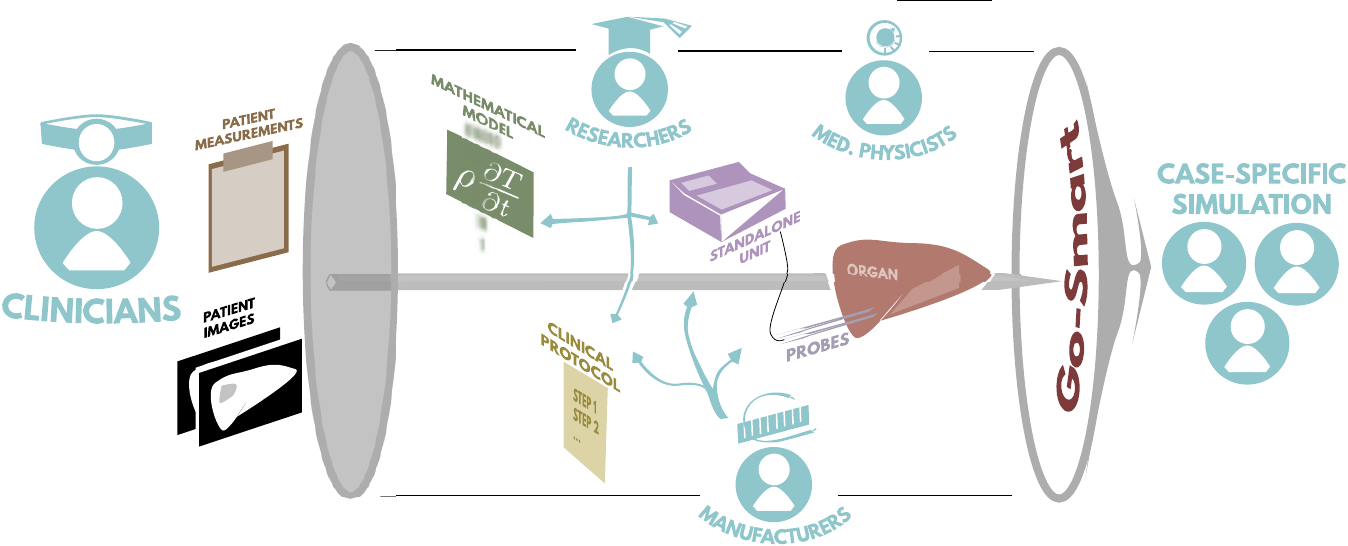}}

  \caption{Graphical abstract}
\end{figure}

\section{Introduction}\label{sec-introduction}

Computer simulation technology can benefit clinicians and patients through
training, education and planning {\cite{krummel1998surgical}}. For greatest
effect, however, direct co-operation between clinicians, engineers and
computational scientists is required {\cite{krummel1998surgical}}. This paper
presents a system of pre-packaged ancillary tools, including a web-based
radiological interface, image manipulation and a simulation architecture. Such
an approach reduces the required collaboration skill-set to those directly
relevant to the research or training at hand, and can remove web development
as a limiting factor in web-based clinical training, commercial product
development and simulation research.

In surgical training, acquisition of basic skills is moving from the operating
theatre to educational settings, building an increasing need for testable,
extensible educational surgical software. Moreover, experienced clinicians
must gain experience using new technology as it is introduced, another
important application of computational and non-computational simulation
{\cite{de2012simulation}}. In general, predicting the outcome of minimally
invasive therapies, particularly those involving thermal ablation, remains
challenging due to the many factors influencing the procedure. Providing a
platform for such training in a specific discipline requires engagement of
both junior and senior clinicians for these reasons.

Manufacturers must also be involved: their consideration of clinicians' needs
is crucial, and includes areas potentially outside core competencies, such as
software user experience {\cite{sawyer1996design}}. This can create a
significant cost and personnel burden in a green-field project. Reduction of
newly designed interfaces can help avoid unpredicted challenges, benefiting
from both user familiarity and a history of integrated feedback. By engaging
with established tools, manufacturers gain insights into clinician and
researcher behaviour at a preliminary stage, prior to costly investment in a
software development or support infrastructure. Structured engagement of
manufacturers with practitioners in simulated settings is encouraged
{\cite{sawyer1996design}}.

The translational link between researchers and clinicians is also an essential
element to support: the existence of a marked gap between research outcomes
and implementation is well-established {\cite{haines2004bridging}}.
Encouraging awareness of developing best practice and technical evolution is
therefore an important role of those engaged in establishing it
{\cite{haines2004bridging}}. Within clinical research itself, the need for
research-relevant information systems has been identified as one of a number
of research challenges, as are infrastructure investment and training. Public
participation, also, is essential and can be impeded by lack of
standardization and conflicts of interest. Together, these emphasize the need
for vendor-independent, research-supporting IT frameworks, complementing
clinical skill-sets and facilitating effective interaction between the worlds
of research and practice {\cite{sun2003central}}.

The importance of effective and well-documented clinical trials is emphasized
by another professional communication gap - that between trial outcomes and
other forms of predictive study. Such discrepancies may arise from publication
bias, inaccessibility of animal study registers and methodological
limitations, amongst other factors {\cite{perel2007comparison}}. A closer loop
must be established between clinical triallists and experimentalists to ensure
shortcomings in preliminary research are identified and resolved
{\cite{perel2007comparison}}. Providing a simplified, standardized and
accessible online interface reduces the practical costs and obstacles to
setting up a clinical or pre-clinical trial, and to the cataloguing and
sharing of trial results.

In contrast to offline, non-web-based systems, an online web-based interface allows
users in different institutions to interact on the same system, and, in particular,
clinicians to use parameters and models dynamically defined by manufacturers and
researchers. It also allows users to access data on their own devices off-site,
for example, when meeting. This is practical only with a web-based system.

The project discussed in this work, the Go-Smart project {\cite{weir2015go}},
provides a web platform for highly adaptable simulation of minimally invasive
cancer treatments. This avoids a computational system that represents a frozen
point of development, allowing new equipment, clinical protocols and numerical
models to be incorporated.

Underpinning this functionality is a domain model relating interventional room
practice, manufacturer equipment and numerical analysis. It provides a
framework for defining allowable combinations of entities, with a hierarchy of
simulation parameters contributed by each. This is exported as a single,
condensed configuration file for the simulation server. Results are viewed
through an interactive, radiology-style web interface and may be validated
against a segmented lesion, if available.

Clinicians interact with the system via a simple point-and-click interface in
the web-site, when launching a simulation through the radiological patient
management system. Entities may indicate certain parameters to be
clinician-specifiable (e.g. duration of optional intervention steps, needle
location), and the clinician is presented with a dynamically generated
web-form. Technical users may access an additional interface, providing
lower-level control of the parameters and properties of their contributed
entities.

\subsection{Novelty}

This sytem is novel in several key respects:

\begin{itemizeminus}
  \item the approach to modelling general medical simulation problems, both
    \begin{itemizeminus}
        \item technically, by combining clinician-input data with researchers'
          sand-boxed, uploaded computer code and parameters, through the Glossia
          system, described below, and
        \item conceptually, through the Clinical Domain Model, described below,
          which structures this approach.
    \end{itemizeminus}
    Together these allow a far greater depth of interaction between researchers,
    manufacturers and clinicians on a single platform than previously possible;
  \item 2D and 3D web visualization techniques, enabling a highly-responsive,
    easy-to-use, web-based radiology platform, comparable to offline systems,
    required for real-world clinician engagement;
  \item expansion and tuning of web-based image segmentation and registration
    techniques to ensure consistent and reliable processing of organ-level
    geometry, required for real-world clinician engagement;
  \item building a range of existing, and novel, numerical models surrounding
    a single medical application - Minimally Invasive Cancer Treatment (MICT) -
    and demonstrating the effectiveness of this system, taking individual patient
    datasets uploaded and segmeneted by clinicians through to simulation, and
    quantified model verification.
\end{itemizeminus}

The Minimally Invasive Cancer Treatment simulation within the Go-Smart project is
dependent on the innovations above, and the results that particular system can
demonstrate have been shown by project authors.

However, the underlying technical and conceptual approach, applicable to general
clinical problems, is first described in detail here. Its outcomes in the MICT
application are shown to demonstrate this system's effectiveness in enabling
highly-effective, clinician-driven use of dynamically-defined simulation.

In the paper, we show a novel architecture that makes online collaboration between
researchers, manufactureres and clinicians possible, without the modelling restrictions
placed on researchers by existing systems.

\subsection{Application}

The specific application under consideration is the web-based analysis of
Minimally Invasive Cancer Treatments (MICTs). MICTs, in the scope of the
project, comprise a set of methods for percutaneous tumor ablations under
image guidance. In this context, we have developed a workflow across a
consortium of nine partners, four clinical and five technical, that has been
tested using incoming patient data over several years. In addition, we have
been supported in our testing by input and feedback from external clinical
users, MICT treatment manufacturers and, as the framework design approach is
intended to be more broadly applicable, external biophysics researchers
operating outside the MICT domain.

MICTs are a growing set of techniques for ablating cancerous tumours without
the need for full surgical resection or when there are no surgical options.
Such techniques include radiofrequency ablation (RFA), microwave ablation
(MWA), cryoablation and irreversible electroporation (IRE), all of which are
performed using percutaneous needles. The specific mechanisms of action, and
underlying physics, are described in {\textsection}\ref{sec-mict-theory}. As
many of these techniques are becoming established and new approaches are
continually appearing, it is challenging for interventional radiologists to
gain and maintain familiarity with the available equipment and indications.
Nonetheless, clinicians must remain aware of the changing field to provide
optimal patient care - in RFA, experience has been shown to be correlated to
treatment outcomes {\cite{hildebrand2006influence}}. Moreover, they must be
experienced in MICT selection and able to simulate or practice with new tools
prior to clinical use. While modality-specific planning tools exist (e.g.
{\cite{rieder2011gpu,kerbl2012intervention}}), previously, there has been no
effective tool for interventional radiologists (IRs) to predict
patient-specific outcomes of a treatment, allowing for various modalities and
equipment of various manufacturers.

This is a clinician-driven requirement to allow patient-specifci comparison of
approaches to treating a given tumour across multiple treatment modalities, as
comparison within a single modality is an artificial constraint to finding an
optimal treatment option.

The Go-Smart project seeks to rectify this, by providing a platform for
clinicians to upload patient data, including CT and MRI images, then to plan,
compare and validate treatment options. For a complete, ergonomic environment
to be achieved, significant development has been undertaken: this encompasses
image segmentation, image registration, simulation, modelling and
visualization, brought together within a purpose-built scalable web framework.
Validation data, for quantifying the performance of the segmentation,
registration, modelling and simulation aspects, has been provided and reviewed
by a series of clinical partners.

A core feature of the Go-Smart framework is its extensibility. To maintain
pace with the state of the art going forward, additional mathematical models,
simulation codes, equipment and even modalities may be added through the web
interface, by independent researchers, technicians and manufacturers.

It is expected that the Go-Smart environment will provide a tool for
independent evaluation of equipment, training of clinicians, collaboration on
treatment planning and medical research.

Even within existing MICTs, clinician experience is a highly determining
factor for patient outcomes. Boundaries of tissue necrosis are hard to predict
heuristically and medical simulation can help reduce patient exposure to
outcomes that are difficult to predict based clinical experience only.
However, a single platform representing a snapshot of technical development in
a rapidly changing field would quickly lose its relevance, and so the ability
to incorporate new technologies through the web interface is indispensible to
this aim.

\subsection{Background}

In the early 2000s, distributed computing projects began to appear in European
health-care research: NeuroGrid {\cite{geddes2004neurogrid}}, MammoGrid
{\cite{amendolia2003mammogrid}} and GEMSS {\cite{berti2003medical}}, for
instance, all began around this time. Applying principles of distributed
computing in this sphere provided effective use of computational resources
between institutions and opportunities for international collaboration.
Complementing standalone tools that provide medical image and/or simulation
analysis toolchains, such as SimBio {\cite{fingberg2003bio}}, COPHIT
{\cite{bartz2003hybrid}}, BloodSim {\cite{penrose2000fluid}}, euHeart
{\cite{smith2011euheart}}, IMPPACT {\cite{payne-et-al-2011-image}}, or, much
earlier, RAPT, newer platforms exist to bring analysis to a distributed
setting, in some cases adapting existing standalone frameworks
{\cite{benkner2003numerical,simpson2010gimi}}. The extension to a generic
toolchain using scientific middleware is a challenge that has been impeded by
closed extensions in proprietary tools {\cite{simpson2010gimi}}, and
evolutionary developments have moved closer to open, standardized protocols,
incorporating web technologies for improved quality of service
{\cite{engelbrecht2009service}}. A more extensive overview of international
computational projects in biomedicine is provided in
{\cite{kramar2013particularities}} and {\cite{redolfi2009grid}}.

Matching earlier distributed tools to a clinically usable web-based interface
is an additional challenge, and one that has more recently been addressed with
projects such as Aneurist {\cite{patel2007development}} or neuGRID
{\cite{redolfi2009grid}}. Simulation and patient-specific analysis algorithms
may be later extensions to distributed data projects, as in the Health-e-Child
and Sim-e-Child projects, respectively
{\cite{freund2006health,ionasec2010patient,mihalef2011patient}}. Other
possible extensions include the expansion of user-facing web services
{\cite{haitas2012distributed}} and broadened application of simulation tools
through intercontinental amalgamation of data sources
{\cite{stamatakos2011exploiting,johnson2013connecting}}. Several such
extensions may be seen in evolution between projects, adding significant value
to existing tools {\cite{burrowes2013multi}}.

Lessons learned from earlier distributed medical projects include the need for
simple, widely-used standards, quality assurance, community building and good
governance {\cite{viceconti2011policy}}. In particular, open infrastructure,
specifically middleware, is a key component of successful projects
{\cite{amendolia2004mammogrid,amendolia2005deployment,wilkinson2002biomoby,ainsworth2007challenges}}.
Adaptable open standards for defining simulation models, such as CellML,
FieldML and SBML, are also important. However, they are not, in themselves, a
panacea for providing modelling compatibility
{\cite{johnson2013dealing,johnson2013connecting,mckeever2015role}}.

Many of these projects will be focused on medical research, rather than the
clinical context. However, bridging the research-application gap within a
platform provides opportunities to augment clinical decision making with
predictive analysis, support clinical trial management, provide access to
accumulated knowledge and, potentially, enhance patient-clinician
communication {\cite{rossi2011p}}. Web-based services may even allow upload
and analysis of data from home-monitoring systems, simultaneously providing a
platform for patient support, clinician review and computational analysis
{\cite{kafali2013commodity12}}. With external clinicians and patients involved
in platform use, a need arises for effective end-user engagement through
channels such as social media {\cite{sakkalis2012bridging}}. Ultimately, if
such tools, especially predictive simulation tools, are intended to inform
medical procedure, an acceptable clinical workflow incorporating the output
analysis must be formed {\cite{thiel2011assessing}}. In the research setting,
an aim of web-based workflow tools, such as Galaxy {\cite{goecks2010galaxy}},
is to enhance accessibility, reproducibility and transparency.

Simulation workflows can facilitate patient-specific predictive analysis,
based on the patient's medical images and quantitative indicators
{\cite{crozier2016image}}. Moreover, they allow for both analysis and training
to be incorporated into a web platform
{\cite{pappalardo2009immunogrid,kononowicz2014framework}}. However,
incorporating computational modelling into education requires clarity around
validation, documentation, copyright, confidentiality, duration and processing
demands {\cite{kononowicz2014framework}}. Simulation workflows may be
primarily dynamically defined {\cite{oinn2006taverna}} or written in a more
imperative, scripted manner {\cite{ciepiela2008gridspace}}. While software
such as Apache Taverna {\cite{hull2006taverna}} and Galaxy
{\cite{goecks2010galaxy}} may provide very flexible workflow definition
toolkits, a need for toolkit interchangeability still exists
{\cite{sakkalis2014web}}. However, an established pattern for a number of
medical simulation applications is the interlinking of an online knowledge
database to a toolchain joining pre-processing, model creation, numerical
solution and post-processing {\cite{bellos2013sifem}}.

Other considerations in recent projects include convenience (no installation
required for end-users), transparency, lightweight components, scalability and
maintainability {\cite{sherif2015cbrain}}. Adaptability, facilitated through
lightweight tooling and a tight developer-stakeholder loop, is essential to
account for the inevitable changeability of requirements throughout a clinical
computing project {\cite{hartswood2005working}}. Where distributed biomedical
projects had generally been grid-based, cloud infrastructure is increasingly
supported, for example, in the CBRAIN and VPH-Share projects
{\cite{sherif2015cbrain,benkner2013cloud}}. In particular, this allows
extension of existing grid projects to a broader audience
{\cite{benkner2013cloud}}. Nonetheless, as cloud resources are often run by
third-party infrastructure as a service (IaaS) providers, quantitative
comparison of cost and performance must be undertaken
{\cite{bubak2013evaluation}}. Within the Go-Smart project, simulation tools
are consequently designed to be flexible with respect to deployment context.

While many of the existing projects have used an administrator-determined or
architect-determined workflow, web-based biomedical simulation design is a
growing approach. This is facilitated by a growth of workflow management
systems, which can provide a convenient ready-made back-end. Such systems
include the early Discovery Net, one of the first such systems, Triana, Apache
Taverna and Kepler {\cite{curcin2008scientific}}. While many such
bioengineering, bioinformatics and biomedical workflow systems now exist, it
is less common to see such systems with an integrated web-based interface for
dynamic inclusion of third-party workflows. Certain projects have based
extended existing workflow packages to improve cloud functionality or
user-friendly web support, such as the Tavaxy project
{\cite{abouelhoda2012tavaxy}}, combining the Galaxy system with the Taverna
suite, WorkWays providing a web gateway for Kepler
{\cite{nguyen2015workways}}, or GPFlow wrapping Microsoft BizTalk and Human
Workflow Services in a user-friendly scientific web interface
{\cite{rygg2008gpflow}}. In an offline setting, the GIMIAS open source
framework provides a set of offline analysis and visualization tools that may
be used inconjunction with Taverna to build complete research workflows, or
preliminary clinical workflows {\cite{larrabide2009gimias}}.

The CHIC project {\cite{stamatakos2014computational,tartarini2014vph}}
provides a general, web-based cancer multi-scale modelling tool. They provide
strong support for model adaptation and flexible architecture, supported by
Apache Taverna workflow management. Moreover, this project incorporates human
intervention steps in model execution and support probabilistic variables.
They define the concept of hypermodels, which may be created by expert users
as a workflow combination of pre-defined hypomodels through the web-based
interface. Clinicians may execute upload patient data and perform simulations
through this interface. While this project forms an important counterpoint to
Go-Smart, the focus in the latter project is towards integrating manufacturer
use cases and basic research with clinical use, by allowing the underlying
hypomodels themselves to be created and managed through the interface. The
CHIC project builds on the broad adaptability of its toolchain, allowing
individual installations to provide new features. In the Go-Smart case, the
clinical domain model, described in
{\textsection}\ref{sec-clinical-domain-model}, is used to provide an
opinionated framework, centred around medical interventions, but facilitating
the cooperative building of new simulation strategies in that context.

The VPH-Share project incorporates a number of disparate medical areas in its
flagship workflows: @neurIST, euHeart, VPHOP and Virolab. Online, or offline,
workflow composition is available through Taverna support
{\cite{testi2014d65}}. Researchers may access a diverse range of components,
and a strong cloud focus provides flexible capacity for computation and data
storage {\cite{koulouzis2013cloud}}. Interactive workflow steps are enabled
through launching interactive interfaces, such as GIMIAS, in a cloud-hosted
virtual machine and providing web-based remote desktop access. This enhances
existing desktop applications through web accessibility. Workflows may be
developed through online tools such as OnlineHPC, or offline with Taverna.
VPH-Share is focused on providing a very general platform supporting diverse
research-driven workflows. In contrast, Go-Smart focuses on a
radiologically-driven design, helping researchers bridge the discipline gap by
adapting and extending the analysis workflow at a low level through scripting,
or a high level through parameter adjustment and manipulation of conceptual
clinical components ({\textsection}\ref{sec-clinical-domain-model}). This
ensures a comfortable, familiar experience for clinicians, while maintaining
flexibility and a low barrier for entry both to developers and high-level
modellers.

Combining Go-Smart utilities with workflow management tools such as Taverna or
OnlineHPC (http://onlinehpc.com), or a web-based research and data management
tool such as VPH-Share, would provide useful future enhancements, integrating
benefits of each project.

\section{Methods}\label{sec-methods}

The Go-Smart distributed architecture is designed to ensure stability and
scalability of the environment. Certain computationally intensive components
have restrictive hardware or software requirements, so processing must be
spread between multiple hosts. The architecture has been divided into
distinct, independently-functioning components with real-time communication
middleware linking physical machines. Clearly-defined interfaces describe
input and output of each component and for the browser and VisApp clients.

\

\begin{figure}[h]
  \resizebox{340px}{!}{\includegraphics{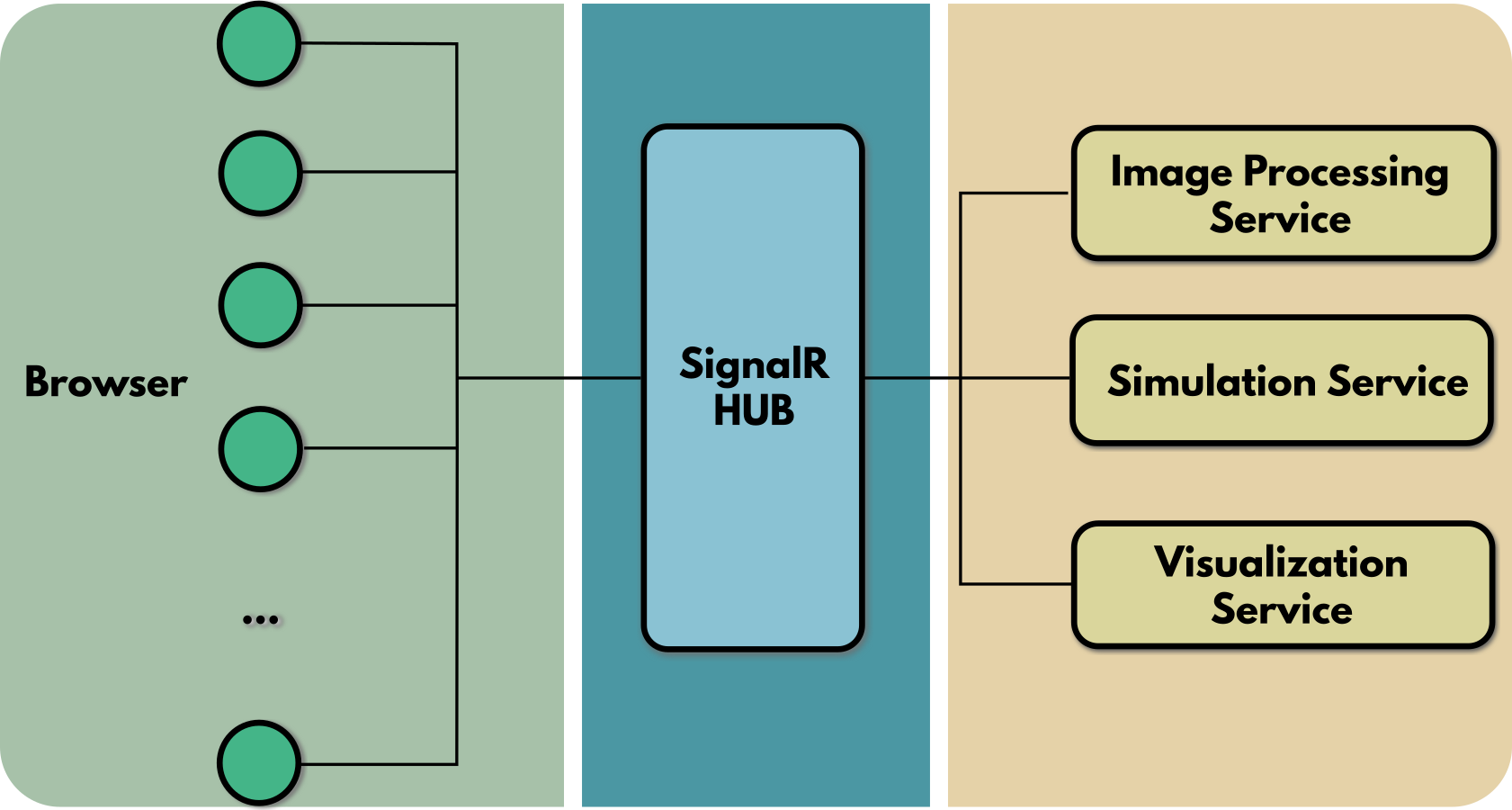}}
  
  \
  \caption{\label{fig-framework-layout}Overview of web application
  architecture. On the far left, various clients connect to the webserver
  running a SignalR Hub that, in turn, connects to simulation, segmentation
  and image processing services elsewhere within a privileged network.
  Business logic and views are provided by a standard ASP.NET MVC application
  hosted on the webserver.}
\end{figure}

\

The user roles discussed in {\textsection}\ref{sec-introduction} have highly
contrasting requirements. Manufacturers want to add new equipment and
algorithms easily. Radiologists want to use this equipment in a simulation
environment, to familiarize themselves with a product, provide a training
environment or to plan an intervention. Finally, researchers validate
simulation results against laboratory tests or real-world clinical data,
compare theoretical models, or experiment with protocol or equipment
formulations.

To address these requirements, the following autonomous services have been
implemented:

\paragraph{Image visualization server}Due to the computational load of certain
computational tasks required to support the radiological interface, a separate
server handles these independently. This service processes the DICOM,
segmentation and simulation data, generating axial, sagittal and coronal
pictures which can be used in any common browser.

In particular, image re-slicing, allowing a user to adjust the axes along
which the coronal, sagittal and axial views are defined, is a computationally
demanding task to be completed with minimal interruption to the user
experience during needle placement. Other tasks related to user-facing image
generation and editing are implemented here, for example, contrast setting and
editing of existing segmentations.

\paragraph{Image processing server}This service segments structures such as
organs, bronchi, vessel trees or tumours in the DICOM data provided by
radiologists. This functionality is described in greater detail in
{\textsection}\ref{sec-image-seg-res}.

\paragraph{Simulation server}The simulation service uses the segmented
structure and further information from parameters or needle positions to
calculate a lesion. As a highly adaptable and flexible series of multi-process
numerical workflows, the simulation framework is a distinct, re-usable
concern, providing simple API endpoints for the web application via the
Crossbar.io middleware. This functionality described in greater detail in
{\textsection}\ref{sec-simulation-orchestration}.

\begin{figure}[h]
  \begin{tabular}{cc}
    \resizebox{340px}{!}{\includegraphics{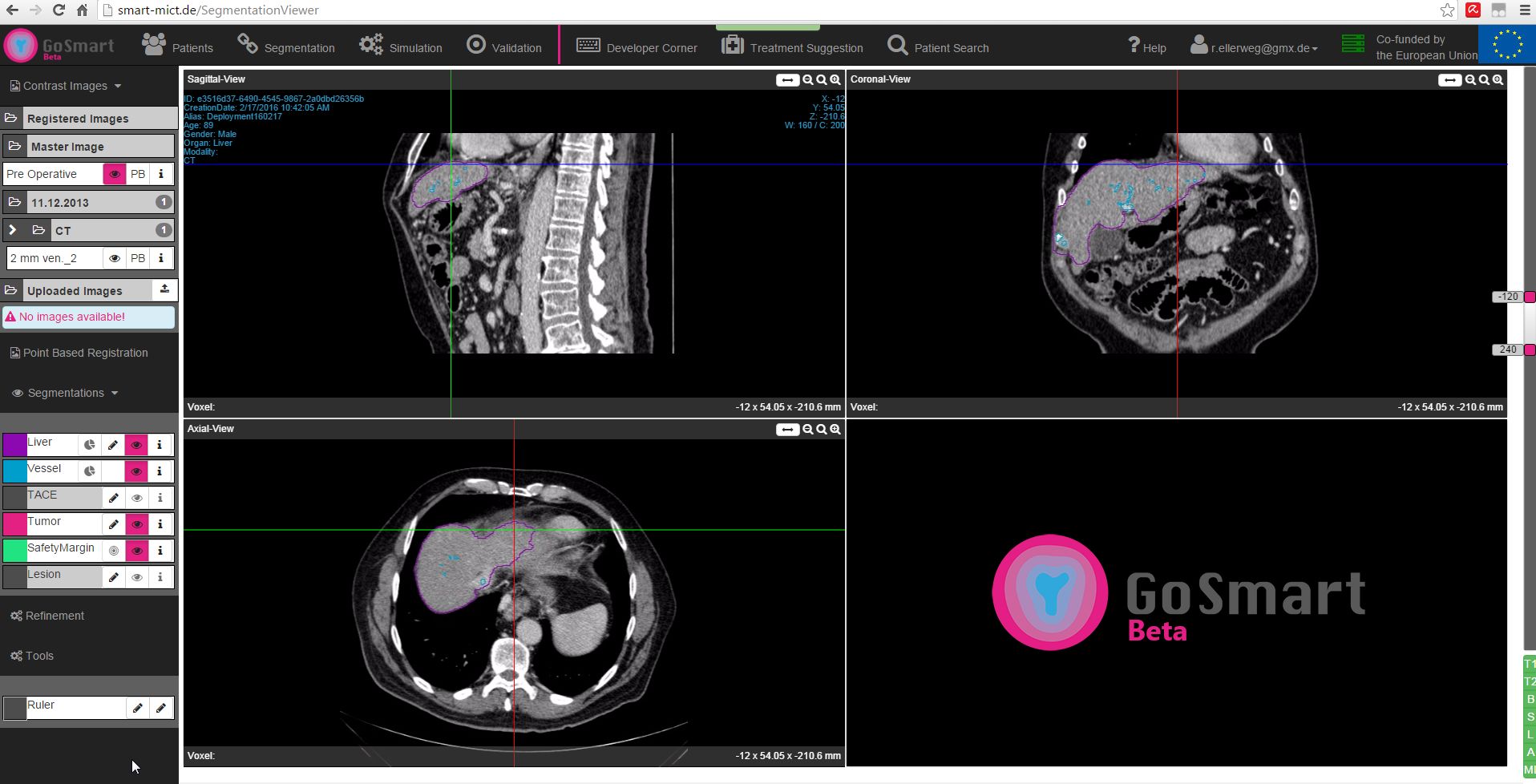}} \\
    (a) Segmentation View \\
    \resizebox{340px}{!}{\includegraphics{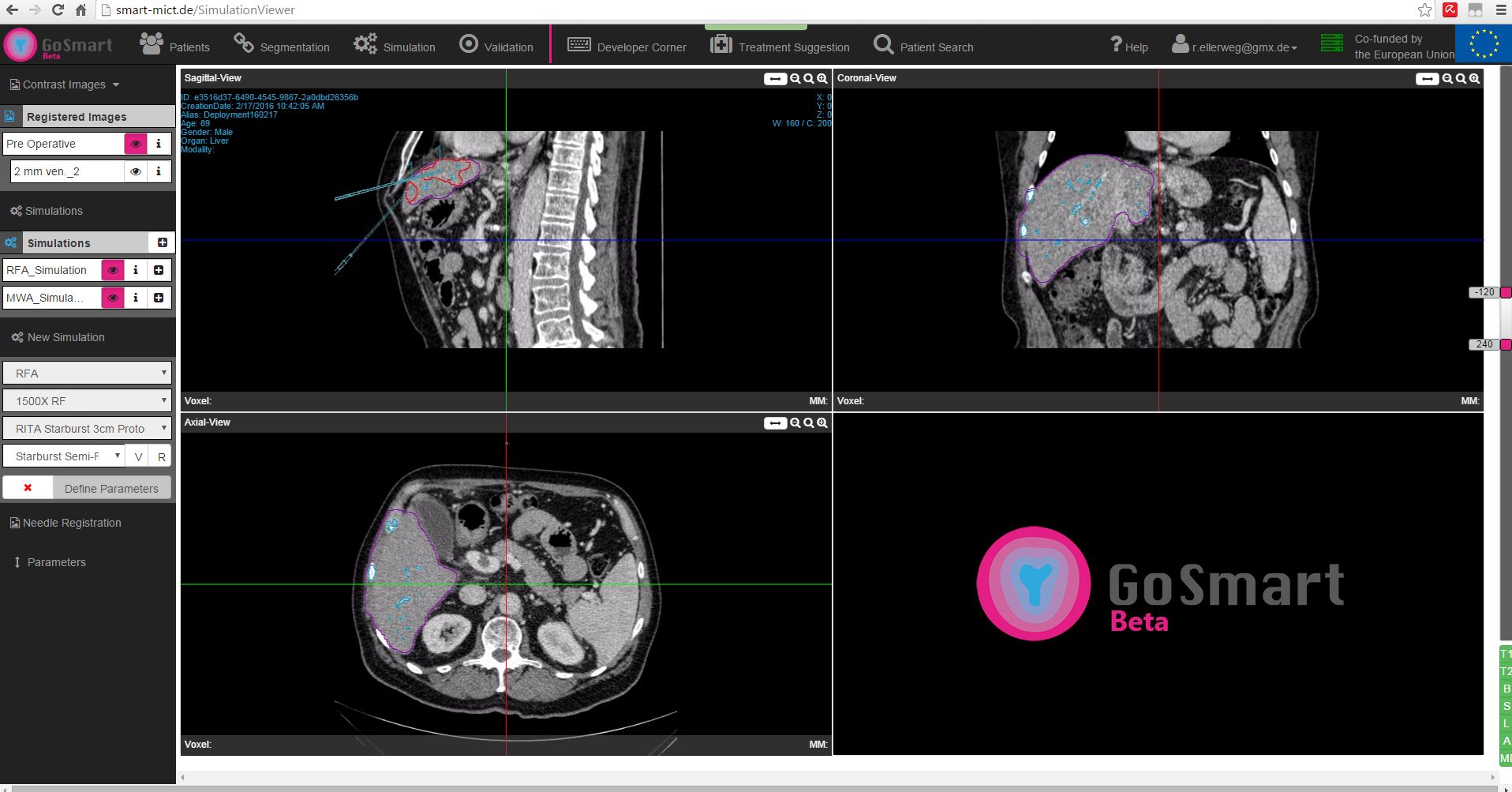}}\\
    (b) Simulation Preparation
  \end{tabular}
  \caption{\label{fig-screenshots-seg-sim}Screenshots of workflow using
  in-browser client. Browser used is Google Chrome}
\end{figure}

\

On the client side, two complementary tools are available:

\paragraph{Browser client}To use the Go-smart environment, the user needs only
a normal web browser. Consequently, there are no special restrictions on the
client machine. This is designed to provide interventional radiologists with a
classic radiology workspace. Figure \ref{fig-screenshots-seg-sim} shows this
layout for segmentation [\ref{fig-screenshots-seg-sim}(a)] and simulation
[\ref{fig-screenshots-seg-sim}(b)].

\paragraph{VisApp client}When a clinician wishes to incorporate 3D interaction
within the interface, they may use a specialized desktop application that
wraps the website and injects a high-quality visualization widget into the
fourth window, alongside the axial, sagittal and coronal views. Otherwise, the
interface remains the same.

\

The services fulfill the basic functional requirements of the three user
groups. For optimal user experience within the web environment, a good
internet connection is required. In particular, the image visualization
service must relay results in real-time, as it should respond efficiently to
brows-er interactions, such as scrolling through a DICOM series, without
visible latency.

The webserver system, hosting an ASP.NET application, marshalls communication
between the client and server-side components. It is responsible for handling
transmission and caching of image data to ensure a favourable trade-off
between data transfer and responsiveness. The application is designed to be
scalable as usage increases. The Go-Smart database is also hosted on the
webserver, which contains the anonymized patient data including medical
images, 3D models and simulation results. To facilitate this, the open source
ASP.NET SignalR library {\cite{netincredibly}} connects services with the
different browser sessions. SignalR uses HTML5 WebSockets to enable
bi-directional communication between clients. If a client cannot use
WebSockets, it falls back to other transport methods such as long-polling.
Figure \ref{fig-framework-layout} shows the discussed components and the
connections between them.

\subsection{Information handling}

\subsubsection{Patient image data and pseudonymization}

Image data is primarily captured in DICOM format, as the standard format for
storing and processing data in medical imaging. The DICOM standard is an
evolving standard, maintained by the DICOM Standards Committee. A
comprehensive standard, it is formatted in 20 sections, covering over 4800
pages. Patient data in a DICOM file is stored with `tags', which may be
referred to using a tag name of the form (NNNN,MMMM) where NNNN and MMMM are
four digit identifiers.

Within a DICOM series of images, no specific metadata file exists, nor a
separate information database for the series. Instead, each image file
contains a header with the metadata content ahead of the image data itself.
Individual manufacturers of image acquisition systems will output DICOM files
with private patient information often stored in non-standard tags.

The Go-Smart project involves four medical partners, who use a variety of
equipment for different treatment modalities from different manufacturers.
Medical image data has to be pseudonymized, with personally-identifiable
references replaced by Unique Identifiers (UIDs), or anonymized before leaving
the originating department. Mappings from UIDs to original cases are stored
securely only at the originating department.

Prior to transfer to an image server, data must be pseudonymized. It is the
responsibility of each department to properly perform this procedure before
using patient data in the web interface. In order to provide a consistent
approach, the following strategy is in place -- all DICOM images used within
the GoSmart project are pseudononymized by using an open source tool,
{\tmem{DICOM Browser}} {\cite{archie2012dicombrowser}}. This tool has the
capability to take a script file with commands for processing individual DICOM
tags and built-in functions for the creation of new UIDs. A consortial
agreement was established that only the following patient information should
be preserved in the pseudonymized data: sex, age, date and time of the
examination.

\subsubsection{Ethics approval}

IRB approval for data collection and processing within the project workflow
was obtained by leading medical partner MUL under local reference AZ
206-13-15072013.

\subsubsection{Case Report Forms}\label{sec-case-report-forms}

All relevant data from patient's history -- both image and clinical data --
are documented within a comprehensive Case Report Form (CRF). This is
available both as an offline document and integrated into the Go-Smart system,
as part of its integrated patient record management. This document contains
general patient statistics, such as age, sex, height and weight. The
preoperative section of the CRF records characteristics of the target tumor
such as size, qualitative perfusion and relevant pre-treatments, alongside
pre-operative imaging notes and relevant blood parameters. During the MICT
procedure, the exact course of the ablation is documented, especially
technical parameters like energy deposition, number of needle positions,
deviations from the protocol and what type of imaging was performed during the
intervention. The post-operative section then evaluates details on follow up
imaging, the size of the treatment lesion and once again relevant blood
parameters.

The CRF can be supplemented by an appendix, which contains screenshots from
key images like target tumor or other relevant imaging information.
Furthermore, adverse events during the treatment must be recorded here.

The use of online CRFs also provides a useful MICT database for future
collaborative research.

\subsection{Usage}

\subsubsection{Primary user groups}

\paragraph{Clinicians}

Once a clinician is logged onto the website, they may add a new patient
record, capturing some or all of the wide variety of fields contained in
standard Case Report Forms ({\textsection}\ref{sec-case-report-forms}).
Pre-interventional CT and/or MRI images may be attached to the patient record
and segmented using a semi-automatic process. Tailored segmentation tools are
available for automatically identifying the lung, liver and kidney as
potential ablation targets (additional organs may be added in future) and
extracting a vessel tree from several contrast phases. With greater clinician
involvement, other visible structures are segmented also, such as tumours and
regions previously targeted by TACE.

Prior to performing a procedure, clinicians may place virtual percutaneous,
needle-like probes in the segmented image, indicating the planned ablation
target (Figure \ref{fig-screenshots-seg-sim}(a)). They set relevant
configuration parameters, perhaps specific to a certain manufacturer or
equipment model, and define an intended treatment protocol. At this point, a
clinician may execute a simulation, which will show the approximate lesion
created by such a treatment (Figure \ref{fig-screenshots-seg-sim}(b)). In most
cases, this unattended simulation process takes 10-60 minutes, although
shorter or longer computations occur for certain combinations of protocol,
equipment and modality.

After performing the procedure, clinicians may upload intra-operative images.
These are then registered to the pre-operative data, which may be produced by
distinct imaging modalities (CT to MRI and vice versa). This allows the
clinician to identify the actual, rather than virtual, location of the
ablation probes in the intra-operative image and so improve the simulation
result.

The simulation models are tuned, by default, to predict the ablation lesion
visible four weeks after the procedure. At this time, the clinician may wish
to upload a set of post-operative images. The actual ablation lesion may be
segmented from this image semi-automatically and, by registration with the
pre-operative image, compared to the predicted outcome. A tool is provided to
quantify the match.

\paragraph{Researchers, developers and manufacturers}

For users who wish to expand the Go-Smart framework, additional workflows are
available. While much of the testing may be done using a workflow similar to
that of a clinician, a technician may use the so-called {\tmem{Developer
Corner}} to add a new mathematical model or piece of equipment.

Numerical models, needles, power generators, organs (contexts) and clinical
protocols all have their own parameters, which may be adjusted through this
interface. Technicians create their own version of a particular piece of
equipment and simulate using their adjusted parameters, or test new
theoretical models or modalities by defining tailored Elmer (finite element
solver) SIF files {\cite{raaback2015overview}} through the Developer Corner.
Support for OpenFOAM {\cite{jasak2007openfoam}} and Python, including the
FEniCS libraries {\cite{logg-2012-automated}} has also been integrated. Due to
the loosely-linked {\tmem{Clinican Domain Model}}
({\textsection}\ref{sec-clinical-domain-model}), re-use of existing or new
numerical models and equipment is simple, allowing a techician to test a new
theoretical model against a range of manufacturers probes, or vice versa.

\subsubsection{Protocols}

For each treatment modality, a tightly or loosely supplied recommendation by
the manufacturer guides clinicians in the execution of the procedure. For RFA,
this is often a complex algorithm, progressing through a series of possibly
repeating steps according to the electrical impedance or average observed
temperature shown on the power generator. For MWA, this is a much freer
decision and clinicians choose a series of powers and durations to perform in
sequence for each location. In cryoablation, rather than power, flow-rate is
varied at set times for each probe. In IRE, probes are set, pairwise, to be
anodes and cathodes with a sequence of potential differences applied between
them. Within all of these protocols a degree of latitude is applied, based on
clinical experience.

As such, the framework maintains a concept of a {\tmem{protocol}}, generally
defined by a technician in Elmer's MATC language {\cite{ruokolainen2015matc}}.
This is normally a feedback loop that relates certain simulation variables to
varying power, or another controlled variable, over time. These protocols
often define the end of the procedure.

To allow straightforward use and customization by clinicians, these protocols
may include parameters set at simulation time, through the user interface, and
may indicate default widgets for clinician interaction (e.g. interactive
power-time-graph).

\subsubsection{Result Analysis}

A separate validation component in the interface allows clinicians and
technicians to view the registered, segmented ablation lesion in the same
interactive pre-operative image viewer as the simulated ablation lesion. Using
the methodology described in {\textsection}\ref{sec-results}, a series of
comparative statistics are calculated and presented.

\subsection{Application components}

\subsubsection{Image segmentation and registration}\label{sec-image-seg-res}

This component allows a patient-specific 3D model to be built for clinician
analysis and simulation, using CT and MRI images, based on a software library
developed during the project. Separate relevant entities, such as organ,
vessels, tumours and observed lesion are identified semi-automatically, with
clinical support and adjustment through the radiological interface.

The target organ may be segmented fully automatically through organ-specific
tools. This is achieved through a rough decomposition of abdominal structures
into semantic objects and morphology-based segmentation.

For cases where fully automatic segmentation is not possible, a series of
`drawing' tools are provided -- these include polygonal, slice-based inclusion
or exclusion of regions; single-click, seed-point segmentation, identifying
contiguous areas; and a free-hand painting tool.

Inner tubular structures, such as vessels, bile ducts and bronchi, are
extracted by an altering Hessian-vessel model based segmentation method
{\cite{alhonnoro2010vessel}}. To provide data for segmenting vessel
structures, a series of images must be obtained at separate phases of contrast
enhancement -- arterial, portal and venous. Segmented structures within these
consecutive images are matched using image registration algorithms, and mapped
into a single frame of references.

Similar techniques are used to match image data from pre-interventional
acquisitions to that of peri-interventional and post-interventional
acquisitions, with users identifying a series of corresponding landmark points
in each image. Pre-interventional to post-interventional registration allows
validation of ablation regions, between simulated and clinician-segmented
profiles, as well as the inclusion of needle positions segmented from
intra-operative imaging in the pre-interventional reference frame, for
simulation.

\subsubsection{Visualization}\label{sec-visapp}

\begin{figure}[h]
  \resizebox{340px}{!}{\includegraphics{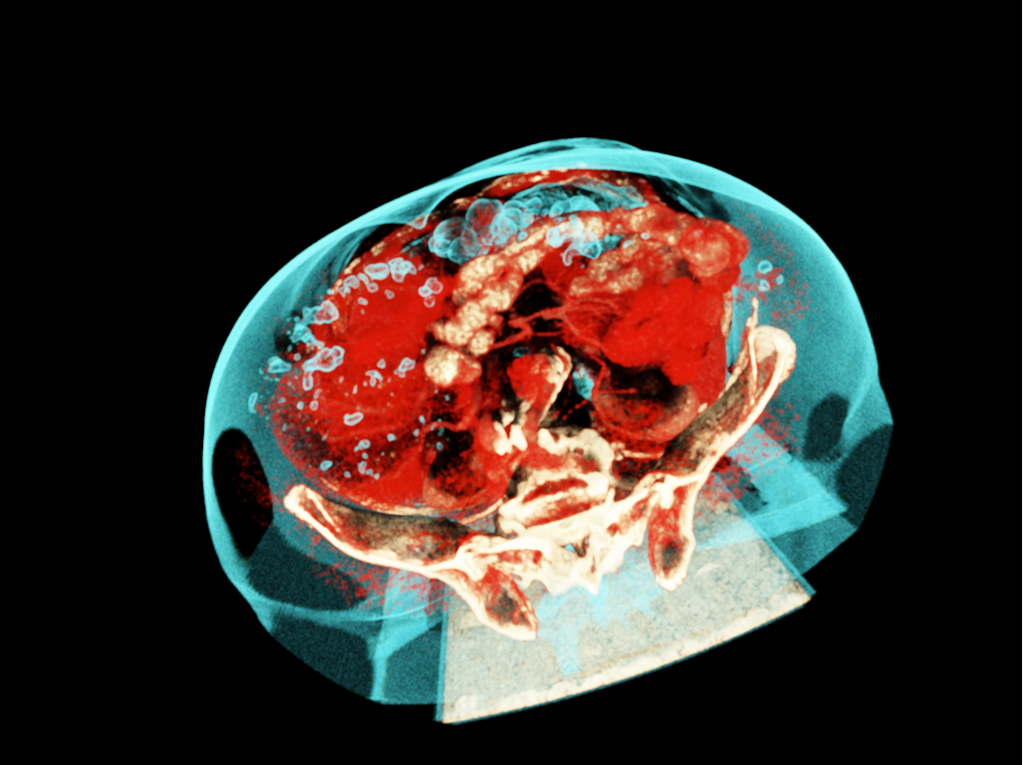}}
  \caption{\label{fig-visapp}Example output of VisApp renderer on high quality
  setting}
\end{figure}

Using the Go-Smart web page in a browser allows a user to visually inspect all
data generated during the workflow in a 2D, slice-based representation. A
standard radiological workstation is provided, encompassing sagittal, coronal
and axial viewers. Within this interface are tools familiar to radiologists,
such as image contrast windowing. Outline profiles of segmentation results,
simulated ablation lesions and percutaneous needles are displayed within these
views.

In addition to this radiological interface, a seamlessly integrated 3D
visualization tool, named VisApp, is available for Microsoft Windows
platforms. It exploits the local computing power of the client PC and
minimizes delays between interaction and response. The application is
implemented as an augmented web browser that identifies a hook on the Go-Smart
web-page, and replaces the corresponding area with a render widget drawing
from local GPU resources. This appears alongside the axial, sagittal and
coronal image windows, providing a fourth window. This view is interactive
and, using the SignalR pathways described in
{\textsection}\ref{sec-methods}, causes the rest of the web-based interface
to be updated in real-time, in response to user actions in the VisApp window
and vice versa.

Advanced 3D volume rendering techniques are employed, which can adapt to the
specifications of the local hardware. From basic direct volume ray casting
{\cite{levoy1990efficient}}, through advanced algorithms with optimal resource
exploitation {\cite{voglreiter2013volume}}, to high-end global illumination
techniques {\cite{khlebnikov-2014-parallel}}, a variety of options are
supplied to the end-user, based on the capacity of the client machine (Figure
\ref{fig-visapp}). Both volumetric data and surface-based representations of
segmented and simulated results, as well as needle models, are visualized
through this interface. This allows user to explore treatment possibilities
and evaluate validation data in 3D, as a supplement to traditional slice-based
techniques.

\subsection{Clinical domain model}\label{sec-clinical-domain-model}

\subsubsection{Outline}

This concept allows a database-persisted set of entities to collectively
define a simulation strategy. The approach taken is more opinionated than
existing model definition approaches (e.g. CellML, FieldML, SGML), in that it
admits only certain, generic types of structure and entity, but allows for the
model definition itself to be an arbitrary parameterized payload. As such, the
clinical domain approach allows us to wrap CellML, say, as a numerical model
taking parameters supplied by one or more clinical domain entities.

\begin{figure}[h]
  \resizebox{340px}{!}{\includegraphics{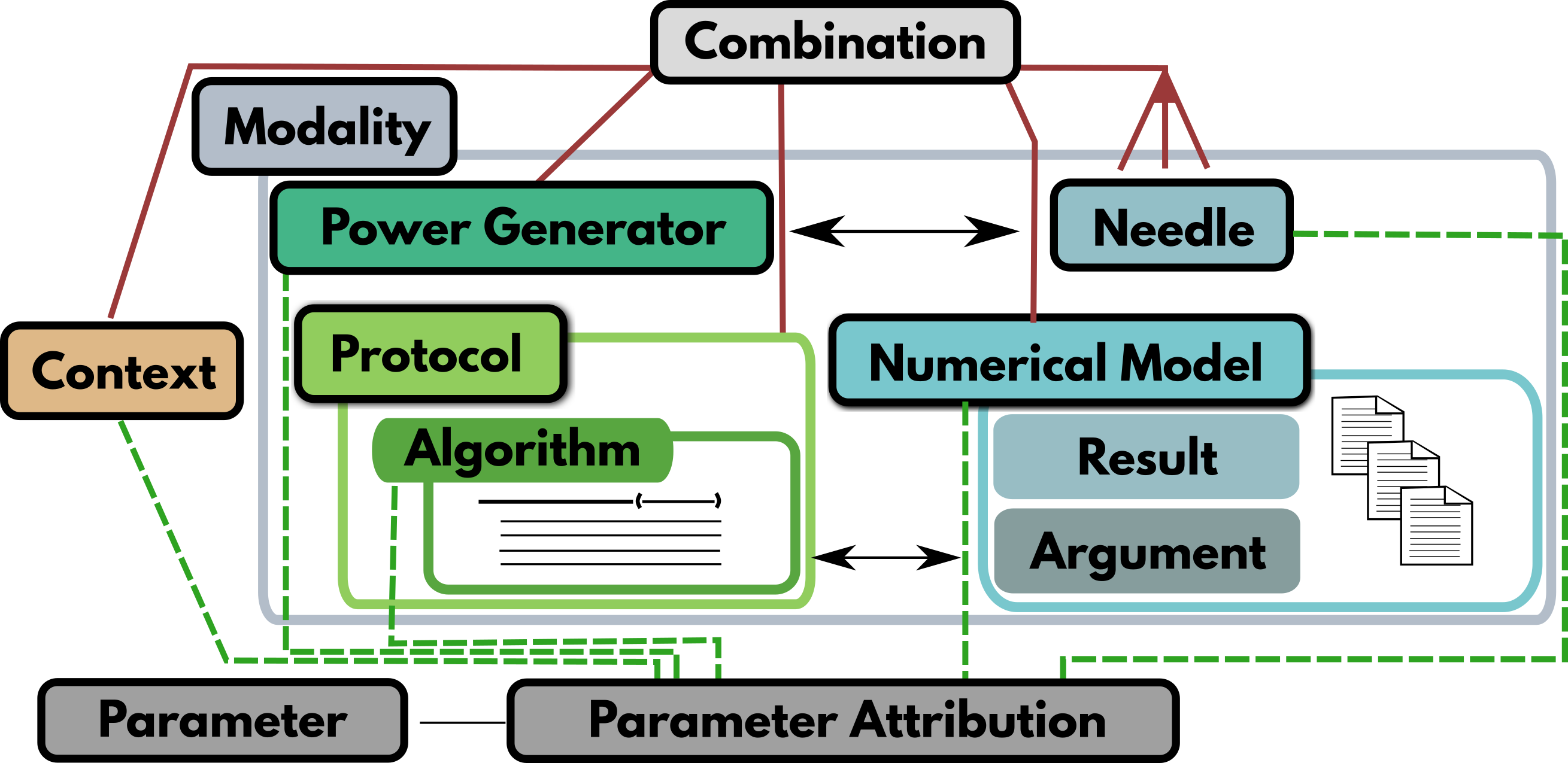}}
  \caption{\label{fig-clinical-domain-relationships}Overview of relationships
  between fundamental entities. Red solid lines indicate component entities of
  a combination (needles with multiplicity), green dashed lines indicate
  parameterizable entities. White lines indicate (usually textual) definitions
  stored in the persistence layer, that is within the entity's record.}
\end{figure}

The fundamental abstract entities are:
\begin{itemizeminus}
  \item {\tmem{Context}}*: environmental parameter set, such as per-organ
  default volumetric constants
  
  \item {\tmem{Power Generator}}*: or any one-per-procedure parameter set. In
  the MICT context, a power generator, regulator or other standalone governing
  machinery. Generally dynamically defined by a manufacturer
  
  \item {\tmem{Needle}}*: or any multiple-per-procedure parameter set. In the
  MICT context, generally a specific manufacturer-model of a percutaneous
  needle. Generally dynamically defined by a manufacturer
  
  \item {\tmem{Protocol}}: a set of algorithms mapping intraprocedural
  variables to model inputs. In the MICT context, generally a clinical
  protocol for indicating recommended clinician interaction with the apparatus
  as the intervention procedes. Generally dynamically defined by a
  manufacturer or clinical researcher
  
  \item {\tmem{Algorithm}}*: a single function definition, with text body,
  that may be interpreted by a particular simulation tool as mapping
  simulation-time inputs to quantitative outputs. In the MICT context, a
  specific output algorithm for a clinical protocol, giving, for instance,
  adjusted power or protocol phase number. Generally dynamically defined by a
  manufacturer or clinical researcher
  
  \item {\tmem{Numerical Model}}*: parameterized definition of numerical model
  or settings for simulation software. In the MICT context, a parameterized
  Elmer (or other FE/FV solver) simulation definition. Generally dynamically
  defined by a medical physicist or biophysical researcher
  
  \item {\tmem{Combination}}: valid, simulatable set of the above entities.
  Generally dynamically defined by a manufacturer or biophysical researcher
  
  \item {\tmem{Modality}}: grouping of above entities, except Context,
  indicating a type of treatment. In the MICT context, this is RFA, MWA,
  Cryoablation, IRE (or other simulatable treatment)
  
  \item {\tmem{Parameter}}: uniquely-named, reusable token representing an
  item of information, possibly with a default value, type and/or entry
  widget. Parameter names are generally in capitalized snake case
  (underscore-separated) for flexibility and ease of identification. This
  appears as, for example, CONSTANT\_INPUT\_POWER.
  
  \item {\tmem{Parameter Attribution}}: a specific application of a Parameter
  to one or more of the starred entities, possibly with overriding value, type
  and/or entry widget
  
  \item {\tmem{Argument}}: a simulatable, intraprocedural output of the
  intervention, such as time or, where appropriate, power, clinical protocol
  phase, temperature observered by apparatus, etc.
  
  \item {\tmem{Result}}: a simulatable output of the procedure
\end{itemizeminus}
The relationship between these entities is depicted in Figure
\ref{fig-clinical-domain-relationships}. In addition to the Combination and
Parameter Attribution relationships noted above, a many-to-many relationship
can specify physically valid linkages of Power Generator and Needle. This
helps a manufacturer to sensibly constrain the possible Combinations that may
later be added by a clinical researcher or other downstream user.

\begin{figure}[h]
  \resizebox{220px}{!}{\includegraphics{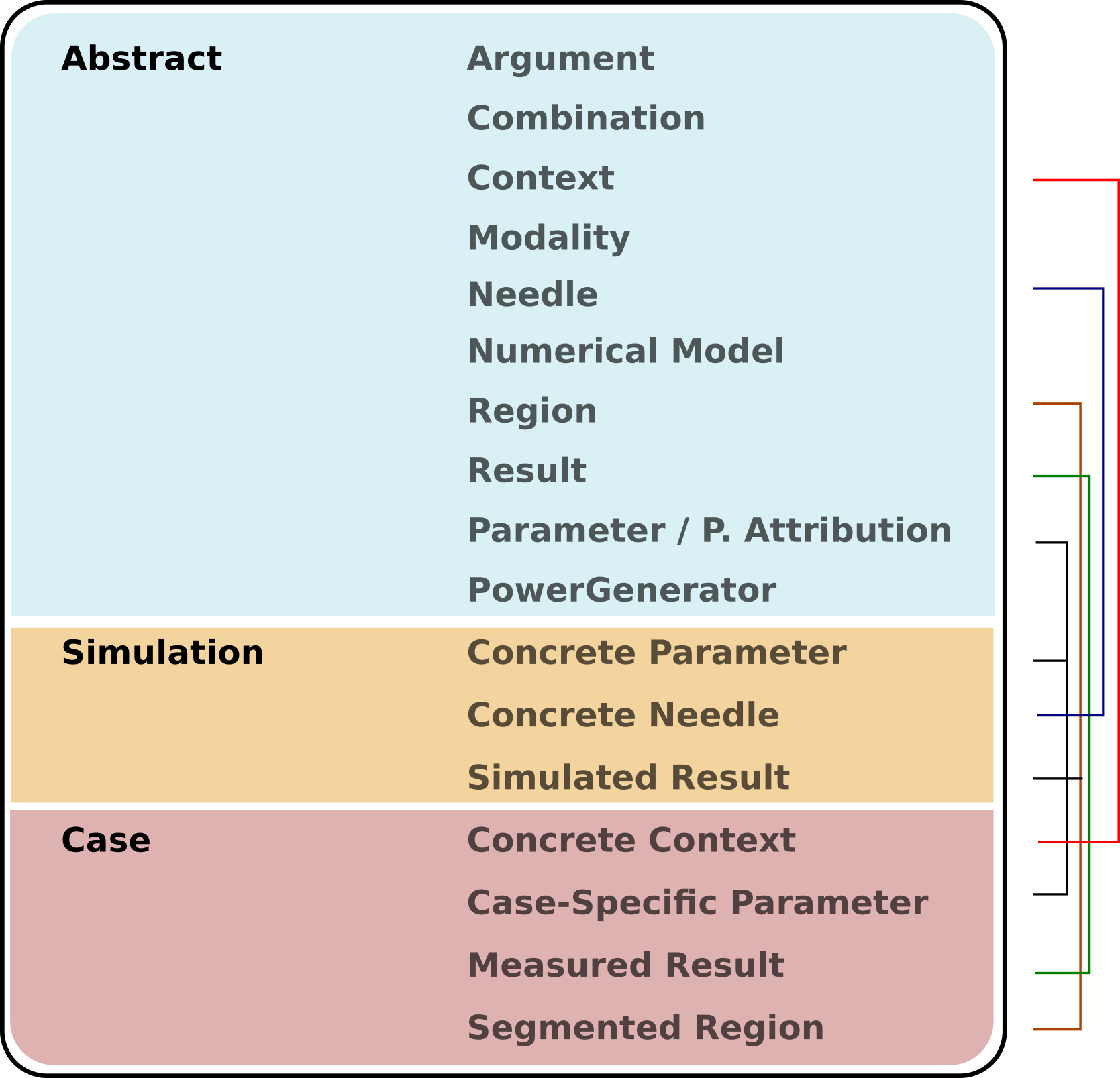}}
  \caption{\label{clinical-domain-tiered}Clinical domain entities arranged in
  Abstract-Simulation-Case (ASC) tiers. Concretization (or measurement)
  relationships are indicated by bracket lines on far right.}
\end{figure}

These entities are stratified into several layers representing abstract
entities, simulation-time entities and case-specific entities (Figure
\ref{clinical-domain-tiered}). Certain of these entities are related through
concretization or measurement, that is, such entities are derived from
abstract forms supplemented by additional information, such as processed image
data or clinician-selected options.

A natural progression may be seen between each tier. Abstract entities are
defined by researchers and manufacturers. Through interaction with the
radiology-style web interface and guided parameter entry, clinicians (or other
users) create a set of simulation entities. This provides a complete set of
data necessary for executing a simulation. Many simulations may be run by
various users against a single Combination -- a set of abstract definitions
representing a procedure -- each simulation creating a new set of Simulation
entities.

For instance, an abstract needle may have a manufacturer model number and a
range of settings. However, until it is {\tmem{concretized}}, that is, used in
a simulation or actual intervention, it does not have a spatial position or a
given number of separately configured needles. A Combination may have several
possible needle model numbers that may be used. Each abstract needle will
represent a different manufacture model. The distinction may be considered
analogous to classes and objects.

A Simulation may have only one of those needle model numbers used several
times, as separate Concrete Needles inheriting from a single Abstract Needle
but each with separate tip and entry locations. Similarly, a Simulation may
have any other repeated selection from the needle types allowed by the
Combination. In the Power Generator -- Needle cross-over table, a manufacturer
may specify the minimum and maximum number of needles that may be used with a
single generator, and so for a given simulation using that Power Generator. In
a more general sense, this provides a separate relationship limiting the
allowed combinations of one-per-simulation entity and multiple-per-simulation
entities.

The third tier represents a set of physical facts relating to the specific
intervention and/or patient, such as patient-specific measurements, image
analysis and intervention outcomes. Some of this data may be used to create a
Simulation, such as the segmented surfaces, or to validate it, such as
measured outcomes.

\subsubsection{Interaction of entities and workflow}

An example workflow is presented showing the creation of entities defined
above in normal practice. For simplicity, an example is chosen where the
manufacturer defines exactly and only the equipment, a researcher prepares a
protocol and numerical model, and a clinician runs a simulation based on their
definitions.

\paragraph{Abstract Tier: Creating Combinations}

\subparagraph{Manufacturer}

When a manufacturer creates a Needle or Power Generator, a new entity in the
Abstract tier is created. The manufacturer defines its characteristics through
Parameters -- these may be additionally constrained to vary by organ
(Context), say, or when paired with an existing clinical protocol (Protocol).
In the case of a Needle, a geometric definition may also be provided.
Parameter Attributions link equipment entities to each Parameter -- if a
Parameter does not exist with that name, one is created.

For each Parameter required, they may indicate whether a clinician can (or
should) fill in a value, and what form widget to use, or whether a Combination
including their equipment should only be definable if this Parameter has been
given a value by another component of the Combination.

\subparagraph{Researcher}

A researcher may add a numerical model to enable simulation of treatments.
They prepare an entity-agnostic numerical model, for instance, using FEniCS
and other scientific Python libraries. That is, it does not depend on the
choice of Context, Needles, Power Generator or Protocols, but only on
Parameters. They create a JSON file, declaring required Parameters and
providing default values if appropriate. In Python, a helper module exists,
which researchers may use to access Parameter and Region information (Example
\ref{ex-python-snippet}). When using the standard simulation workflows, for
Elmer and FEniCS, researchers can assume a volumetric mesh has been provided
in GMSH MSH format before their code runs, and all volumetric and boundary
subdomains have known indices. The researcher may indicate required regions,
with admissible multiplicities, in their model definition -- these will be
supplied by the output of the segmentation tool-chain.

Parameter values may be any JSON-representable type, although basic types
provide most flexibility.

\begin{example}
  \label{ex-python-snippet}{\tmem{Python snippet based on an IRE Numerical
  Model}}. This converts the CGAL-supplied mesh to DOLFIN format and loads it.
  IRE electrode data is loaded from a Parameter and the mesh indices for
  vessel regions are extracted as integers. These may then be used for
  providing, e.g., boundary constraints and solving using FEniCS.
  
  {\noindent}\begin{indent}
      \ttfamily{from gosmart.parameters import P, R
      
      import dolfin as d
      
      import subprocess
      
      \
      
      subprocess.call([
      
      \ \ \ "dolfin-convert",
      
      \ \ \ "/shared/output/run/input.msh",
      
      \ \ \ destination\_xml
      
      ])
      
      mesh = d.Mesh(destination\_xml)
      
      electrode\_triples = P.CONSTANT\_IRE\_NEEDLEPAIR\_VOLTAGE
      
      vessel\_mesh\_indices = [r.idx for r in R.group('vessels')]}
  \end{indent}{\hspace*{\fill}}{\medskip}
\end{example}

In the Elmer case, using Jinja2-templated SIF files (http://jinja.pocoo.org),
the syntax for Parameter access is similar.

For each Parameter required, a researcher may indicate whether a clinician can
(or should) fill in a value, and what form widget to use, or whether a
Combination including their model should only be definable if this Parameter
has been given a value by another component of the Combination.

\subparagraph{Combination creation}

Either the researcher or manufacturer may add a new Combination, or set of
Combinations, to make use of these models and equipment. All Simulations are
based on an existing Combination.

Each Combination includes a single Numerical Model, Power Generator, Context,
Protocol (which may be trivial, not supplying any algorithms) and one or more
Needles. When a Combination is defined, the system confirms that all
Parameters defined by a member of the Combination are either filled with
default values, or allow values to be provided at simulation-time. Otherwise,
it will refuse to create the Combination -- this provides a crucial validation
step, ensuring equipment and contexts, such as organs or phantoms, are
pairable with the numerical models.

\paragraph{Simulation and Case Tiers: Using Combinations}

\subparagraph{Clinician}

A non-technical user, or technical user following their workflow, will upload
patient data or experimental data. This creates certain Case Tier entities,
such as defining a specific concrete context, indicating that this case is in
a human liver, say, or an {\tmem{ex vivo}} muscle sample. The segmentation
process identifies a series of regions with concrete geometric extents.

The user may then wish to run a Simulation. Only existing Combinations may be
used for Simulation. The user picks a Power Generator and Clinical Protocol
they wish to use via HTML drop-down boxes. None, one or more Needles (as
permitted by the model) are added by the user via another drop-down box. The
user identifies a tip location and organ entry-point for each by clicking in
the radiology planes. These Concrete Needles, a series of Abstract Needles
augmented with a specific geometric location, are saved.

Once the Needles are placed, only the Numerical Model remains a free
selection. For each such selection, the administrator chooses one Combination
to be marked {\tmem{public}}, to avoid non-technical users ever being
confronted with a choice of Numerical Models. However, this public marker does
not apply to technical users, and they are presented with a final drop-down
menu of Numerical Models.

This process specifies the Combination completely, and the associated
Parameters which have been marked for clinician input are presented in a
user-friendly HTML form, alongside the radiology view. When they finally
execute the Simulation, the Parameter values are saved as Concrete Parameters
attached to the Simulation. They are no longer related to their source entity,
except in the case of Needle Parameters, which are then tied to the Concrete
Needles added above.

The Simulation definition is sent to the simulation orchestration tool via
WAMP middleware. In the Go-Smart context, this is a transition from a .NET web
application to one or more Linux back-end hosts.

\subsection{Simulation orchestration}\label{sec-simulation-orchestration}

In creating a new simulation, the synthesis of members of a Combination into a
single set of simulation settings is the responsibility of the web
infrastructure, rather than the simulation orchestration tool. To enable clean
exchange of model settings, without leakage of entity identities, a simple XML
sublanguage has been defined: GSSA-XML. This contains only Simulation tier
information necessary to execute a numerical analysis (i.e. that required by
the Numerical Model entity). Its purpose is to prevent entity-specific
behaviour on the simulation side, and ensure application-side entity
extensibility independent of the simulation tool.

The orchestration tool that converts GSSA-XML into workflow-specific settings
and launches individual simulation containers, is an open source package named
Glossia, developed as part of Go-Smart. It hooks into the middleware by
providing a series of WAMP end-points. While any WAMP router is adequate, a
sample configuration is provided for the Crossbar.io software. This
architecture enables a simulation to be executed and monitored from a process
on any accessible machine in any language that has WAMP bindings. Glossia
interaction has been tested from C\#, Python, PHP and Javascript clients. The
language of the workflow tools, inside the simulation container, is entirely
independent of Glossia.

The Glossia server accepts a GSSA-XML simulation definition and examines it
for a specific Numerical Model {\tmem{family}} -- this is part of the
Numerical Model's entity definition. A family is defined to be a pair: a
specific Docker image for performing numerics and a Python `mortar' module,
used to translate GSSA-XML into the necessary numerical package configuration.
For Python-scripted simulation workflows, which can import the Glossia Python
Container Module, no additional plug-in is required, making the creation of
additional Docker families simpler. The developer may access all needles,
parameters and computational regions through this module as normal Python
structures, leaving communication detail to the Glossia tools.

For a given family, the configuration may be very general. In the case of the
FEniCS {\cite{logg-2012-automated}} family, the configuration is a numerical
Python module that may use any of a range of numerical Python libraries
accessible inside the container, and possibly not the FEniCS library itself.

\subsubsection{Docker simulation containers}

The usage of Docker in non-scientific contexts has developed rapidly over the
last two years. It provides lightweight, walled-off environments, with many of
the benefits and few of the drawbacks of virtual machines. In particular,
performance is similar to what may be seen without containerization. It is in
production, providing user sandboxing for many Platform-as-a-Service (PaaS)
providers. Its use in science and medicine is gradually being explored, as
awareness of and experience in these technologies spreads from the web
development and computer science.

As well as supporting a secure environment to run untrusted code, it allows
reproducability, ensuring that a researcher's code will run on the Glossia
server exactly as it does on the researcher's personal workstation. To
facilitate this, all Glossia tools are provided as open source software
(https://github.com/go-smart), mostly under the Affero Gnu Public License
(AGPL). Example families are included, wrapping the FEniCS and Elmer
{\cite{raaback2015overview}} numerical suites.

The AGPL requires a third party that wishes to provide a Go-Smart-like web
service using Glossia to release any incorporated Glossia modifications,
including addition of `mortar' modules. However, the contents of new
simulation container images or separate software on the host are not affected
by this license and may be kept private (notwithstanding licensing
restrictions of other software components). Libraries for use inside
containers are provided under the MIT license, allowing proprietary or private
code to be used for simulation if desired. The sale of more relaxed licensing
is a potential component of project exploitation.

\section{Theory of Minimally Invasive Cancer
Treatments}\label{sec-mict-theory}

The core mathematical models for simulating a series of ablation treatments
are outlined. All are implemented, for the purposes of the web interface,
using the Elmer framework, although some have purpose-written research
implementations also in Python (FEniCS) or OpenFOAM. In most cases, the
simulation process takes 10 - 60\,min, unattended, for adequately refined
results. However, as the environment permits on-going improvement and
refinement of models, timing statistics are not fixed. While these models are
based on established, published algorithms, additional project research has
refined them and allowed identification of key patient-specific parameters
{\cite{hall-2015-cell}}.

\subsection{Common Models of Thermal Modalities}

Several of the modalities function by using hypo- or hyperthermia to destroy
tissue. Two models, in particular, are thus shared:

\subsubsection{Bioheat equation with perfusion term}

As evidenced by analysis during the IMPPACT project
{\cite{kerbl-et-al-2013-intervention}}, {\cite{payne-et-al-2011-image}}, and
following work of Kr{\"o}ger et al {\cite{kroger-et-al-2006-numerical}}, a
basic Pennes bioheat equation with added perfusion term is adequate to model
first order effects of the thermal modalities. This formulation is used in a
variety of numerical approaches to MICT modelling
{\cite{audigier2015efficient}}. The governing equation is,
\begin{eqnarray*}
  \rho c \partial_t T - k \nabla^2 T & = & Q_{\tmop{inst}} + Q_{\tmop{perf}},
\end{eqnarray*}
where $\rho$, $c$, $k$ and $T$ \ are the density, specific heat capacity, heat
conductivity and temperature of the perfused tissue, respectively.
$Q_{\tmop{inst}}$ represents the heat flux due to the ablation instrument. The
norming effect of tissue perfusion is defined to be,
\begin{eqnarray*}
  Q_{\tmop{perf}} (x) & = & \left\{ \begin{array}{ll}
    \nu \rho_b c_b  (T_{\tmop{body}} - T (x)) & \tmop{if} D (x) \geqslant
    D_0,\\
    0 & \tmop{otherwise}
  \end{array} \right.
\end{eqnarray*}
with $x$ being the spatial coordinate, $D (x)$ indicating the local fraction
of cells considered dead and $D_0$. The perfusion coefficient, $\nu$ is a
material property of the local medium (e.g. lung tissue, liver tissue, tumour
tissue), $T_{\tmop{body}}$ is standard body temperature and $T$ is the current
local temperature. $\rho_b$ and $c_b$ represent the density and specific heat
capacity of blood. Note that perfusion here is taken to be a field value,
representing the thermal effect of blood-flow in vessels smaller than the
imaging resolution.

\subsubsection{Cell death model}

While under hypothermia a simple empirical isotherm is used, for the case of
hyperthermia, a more complex cell death model is used, developed during the
IMPPACT project {\cite{oneill-2011-three}}. Cells exist in one of three
states: Alive, Vulnerable and Dead. Cells may transition from being Alive to
being Vulnerable, from being Vulnerable to being Dead and from being
Vulnerable to being Alive. The rates at which they do so are dependent on
local temperature.

The fraction of cells at a location in an Alive, Vulnerable or Dead state is
expressed in terms of $A$, $V$ and $D$, respectively. Each variable lies
between 0 and 1 and the sum of all three is consistently 1.0 -- this allows
the Vulnerability variable to be removed from the algorithm. The relationships
are expressed as follows,
\begin{eqnarray*}
  \frac{d A}{d t} & = & - \bar{k}_f e^{T / T_k}  (1 - A) A + k_b  (1 - A -
  D),\\
  \frac{d D}{d t} & = & \bar{k}_f e^{T / T_k}  (1 - A)  (1 - A - D),\\
  &  & A|_{t = 0} = 0.99, D|_{t = 0} = 0.0
\end{eqnarray*}
The forward and backward rate coefficients, $\overline{k}_f$ and
$\overline{k}_b$ may be constants or temperature dependent. By default, the
lesion is estimated to be the region in which $D \geqslant D_0 := 0.8$.

\subsection{MICT-Specific Models}

\subsubsection{Microwave ablation}

This modality is modeled by coupling the above bioheat equation and death
equation to a simplified Maxwell's Equations solver. The local value of
$Q_{\tmop{inst}}$ is calculated using a transverse-magnetic (TM) axisymmetric
cylindrical solver. The primary equation is,
\begin{eqnarray}
  \nabla \times \left[ \left( \varepsilon_r - \mathrm{i}  \frac{\sigma}{\omega
  \varepsilon_0} \right)^{- 1} \nabla \times \tmmathbf{H} \right] - \mu_r
  k_0^2 \tmmathbf{H} & = & 0,  \label{eqn-maxwell}
\end{eqnarray}
where $\varepsilon_r$ and $\sigma$ are temperature-dependent relative
permittivity and conductivity of tissue, respectively, at the manufacturer's
stated frequency. $z$ and $r$ are the local cylindrical coordinates along the
probe shaft and centred on its tip. $k_0$ and $\omega$ are the wave number in
a vacuum and angular frequency, respectively. $\tmmathbf{H}= H_{\phi}
\mathe_{\phi}$ is the magnetic field vector, approximated as having only an
azimuthal component, $H_{\phi}$. This may be used to calculate the local
energy deposition into the tissue (SAR), through the relationship,
\begin{eqnarray}
  Q_{\tmop{inst}} & = & \frac{1}{2} \sigma | \nabla \tmmathbf{H} | . 
  \label{eqn-mwa-sar}
\end{eqnarray}
As parameters in Equation \ref{eqn-maxwell} are temperature dependent, we can
see the nonlinear coupling between the bioheat and coax equations. Unlike the
simpler, linear model, without temperature dependent electromagnetic
parameters, the nonlinear coupled problem tends to a steady-state solution
over relatively short timescales. The steady-state magnetic field is obtained
from the axisymmetric model with a fine resolution numerical mesh. This then
supplies $\nabla \tmmathbf{H}$ for Equation \ref{eqn-mwa-sar}, which may be
used with temperature-varying $\sigma$ in a time-stepping bioheat model. This
enables microwave simulation to be performed in clinically applicable times.

The current model is based on the Amica Microsulis prototype probe, although
on-going work is incorporating a second manufacturer's configuration.

\subsubsection{Cryoablation}

While the basic model for this is the modified Pennes bioheat equation, a
front-capturing multi-phase solver is used to ensure accurate representation
of the change in physical properties due to the expanding ice ball. In
particular, the numerical method used is the {\tmem{effective heat capacity
method}}. Here, the latent heat of phase change is accounted for by an
adjustment to the relationship between specific heat capacity and temperature.
Moreover, a {\tmem{mushy}} region is admitted, permitting a smooth transition
of physical properties. Mathematically, the effective heat capacity is given
as,
\begin{eqnarray*}
  c_{\tmop{eff}} (T) & = & \left\{ \begin{array}{ll}
    c_s, & T < T_s,\\
    \frac{c_s + c_l}{2} + \frac{h_{\tmop{sf}}}{2 (T_l - T_s)}, & T_s \leqslant
    T \leqslant T_l,\\
    c_l, & T > T_l,
  \end{array} \right.
\end{eqnarray*}
where $h_{\tmop{sf}}$ is the latent heat of solidification, $T$ is
temperature, $c$ is heat capacity and the subscripts $l$ and $s$ denote the
{\tmem{solidus}} and {\tmem{liquidus}} states, respectively.

The thermal conductivity is then defined to be,
\begin{eqnarray*}
  k_{\tmop{eff}} (T) & = & \left\{ \begin{array}{ll}
    k_s, & T < T_s,\\
    k_s + \frac{1}{2 (T_l - T_s)}  (k_l - k_s)  (T - T_s), & T_s \leqslant T
    \leqslant T_l,\\
    k_l, & T > T_l,
  \end{array} \right.
\end{eqnarray*}
where $k$ is thermal conductivity. These effective values replace their
equivalents in the Pennes bioheat equation and the resulting nonlinear
equation is solved iteratively.

This model and protocol has been tested against ablations performed using
Galil Medical Systems' cryoablation technology.

\subsubsection{Irreversible electroporation}

IRE is modeled using a simple electric potential solver,
\begin{eqnarray*}
  \nabla \cdot (\sigma \nabla \phi) & = & 0,\\
  \frac{\partial \phi}{\partial n} |_{A_i} & = & V_i,\\
  \frac{\partial \phi}{\partial n} |_{C_i} & = & 0,
\end{eqnarray*}
with conductivity $\sigma$ and electric potential $\phi$. $A_i$ and $C_i$ are
the $i^{\tmop{th}}$ anode and cathode, respectively, and $V_i$ is the defined
potential difference between them.

Over the ordered sequence of pairings in the protocol, each of which is
defined by two of up to six probes and their potential difference, the
electrical properties change based on the deposited energy. The final lesion
is defined as an isovolume based on a chosen threshold of the local energy
maximum over the whole protocol sequence
{\cite{pucihar2011equivalent,garcia2014numerical}}. Future work is required to
determine precise parameters for heuristic thresholds.

\subsubsection{Radiofrequency ablation}

Rather than performing a Joule heating simulation for each execution of this
modality, an empirical approach is generally used, consisting of a summation
of Gaussian functions centred on suitably chosen points. This technique was
validated during the IMPPACT project, and avoids the extremely large meshes
required to capture the $< 1 \, \text{mm}$ diameter probe tines. Further
studies using a calculated Joule heating field are planned. This model is
tailored to the Angiodynamics RITA family of RFA needles, although extension
work is on-going to incorporate extensible-tine probes of another
manufacturer.

The RFA power generator simulation accounts for the feedback loop based on the
thermocouples at the tips of individual tines. Recommended clinical protocols
may be followed automatically, where the simulation moves from one protocol
step to the next based on temperature readings, for instance. Tines may be
extended or retracted according to the protocol, and the input power is
governed by a PID controller, with temperature as the primary variable (for
RITA needles).

\subsection{Extension}

Mathematical models applied to other areas of the body or treatment
methodologies may be integrated by creation of finite element configurations
for already-incorporated simulation families, such as Elmer or FEniCS. As
these families are modular, third party tools may be incorporated by the
server administrator, as a separate Docker image, with preprocessing Python
plug-in modules. Numerical models built on top of this new family may be
defined in any way parsable by the plug-in module and container image.

\section{Results \& Discussion}\label{sec-results}

Sample results obtained using the Go-Smart workflow, demonstrating the
effectiveness of the procedure, are presented. In keeping with the focus of
the current work on simulation orchestration and model-building methodology,
full validation studies examining the effectiveness of biophysical models of
each MICT procedure will be the subject of future treatment-focused
communications.

It should be reinforced that, within the context of this paper, our primary
demonstration through these results is the effectiveness of the system for
bringing together parameters, models and patient-specific data into
representative simulated outputs. Full analysis and detailed validation of the
models themselves is provided separately.

\subsection{Evaluation measures}

To provide evaluation of simulation performance, the ablation lesion as
observed 4-6 months after the intervention is used as a reference. This is
segmented by clinicians through the interface and may be registered into the
same coordinate system as the pre-interventional segmentations, on which the
simulation is based.

For demonstrating applicability to end-user workflows, comparison of size and
shape is of primary interest. As organ-based registration between the pre- and
post-interventional image acquisitions is affected by alterations in the
relative locations of internal structures over the 4-6 week intervening
period, a rigid-body registration is applied between the segmented and
simulated ablation lesions. While this prevents us from measuring offset, it
allows us to quantify shape and size deviation between the two ablation zones.
Potential effects of this approach on measuring simulation accuracy are raised
in the discussion of results.

The segmented ablation lesion volume, after rigid-body registration, is
denoted $S$. The simulated ablation lesion volume is denoted $\Sigma$.
Measures presented here are the {\tmem{DICE}} metric [DICE], {\tmem{target
overlap}} or {\tmem{sensitivity}} [SN], {\tmem{positive predictive value}}
[PPV] and {\tmem{average absolute error}} [AAE]. Measures used are established
measures in image comparison
({\cite{klein-2009-evaluation,audigier2015efficient}}). These are defined as
follows:
\begin{eqnarray*}
  \text{DICE} & = & \frac{2 | S \cap \Sigma |}{| S | + | \Sigma |},\\
  \text{SN} & = & \frac{| S \cap \Sigma |}{| S |},\\
  \text{PPV} & = & \frac{| S \cap \Sigma |}{| \Sigma |},\\
  \text{AAE} & = & \int_{\partial S} \{ \inf_{\tmmathbf{y} \in \partial
  \Sigma} | \tmmathbf{x}-\tmmathbf{y} | \} d\tmmathbf{x}/ | \partial S | .
\end{eqnarray*}
The DICE, SN and PPV values range from least match at 0.0 to identical overlap
between $S$ and $\Sigma$ at 1.0. While the DICE value is symmetric, the SN and
PPV together help indicate the dominant type of mismatch -- false positive or
false negative. The AAE value is an average of the minimum distance to a point
on $\Sigma$ from each point on $S$, and helps isolate surface comparison when
considered alongside the volumetric measures. The mean, $\mu$, and standard
deviation, $\sigma$, of each measure is presented. The volumetric ratios are
unitless. AAE is measured in millimetres. All values are rounded to 3d.p.

\

\subsection{Radiofrequency ablation}

\begin{table}[h]
  \begin{tabular}{l|lllll}
    & Organ & DICE & SN & PPV & AAE ($\text{mm}$)\\
    \hline
    RFA1 & Liver & 0.591 & 0.828 & 0.459 & 3.853\\
    RFA2 & Liver & 0.711 & 0.818 & 0.628 & 3.393\\
    RFA3 & Liver & 0.631 & 0.661 & 0.604 & 2.975\\
    RFA4 & Liver & 0.657 & 0.503 & 0.946 & 2.957\\
    RFA5 & Liver & 0.694 & 0.785 & 0.622 & 2.792\\
    \hline
    $\mu$ &  & 0.657 & 0.719 & 0.652 & 3.194\\
    $\sigma$ &  & 0.043 & 0.123 & 0.160 & 0.384
  \end{tabular}
  \caption{\label{tab-eval-rfa}Evaluation measures for a series of RFA
  treatments}
\end{table}

All radiofrequency ablation interventions presented in Table
\ref{tab-eval-rfa} were performed at the Leipzig University Hospital [DE].
Through the web interface, clinicians at this institution uploaded patient
image data, segmented and registered it, then prepared RFA simulations to
match their treatment.

Cases RFA1-RFA5 do not generally show a larger SN value than PPV value or vice
versa, suggesting that the simulated ablation zone is not systematically over-
or under-estimating the segmented ablation zone size. The AAE is steady,
$2.792 \, \text{mm}$ - $3.83 \, \text{mm}$, indicating a fairly consistent
shape deviation. Where the AAE is larger, the PPV is significantly lower than
the SN, suggesting that these higher deviation cases are due to a
overestimation of the true lesion size by the model.

\begin{figure}[h]
  \begin{tabular}{cc}
    \resizebox{160px}{!}{\includegraphics{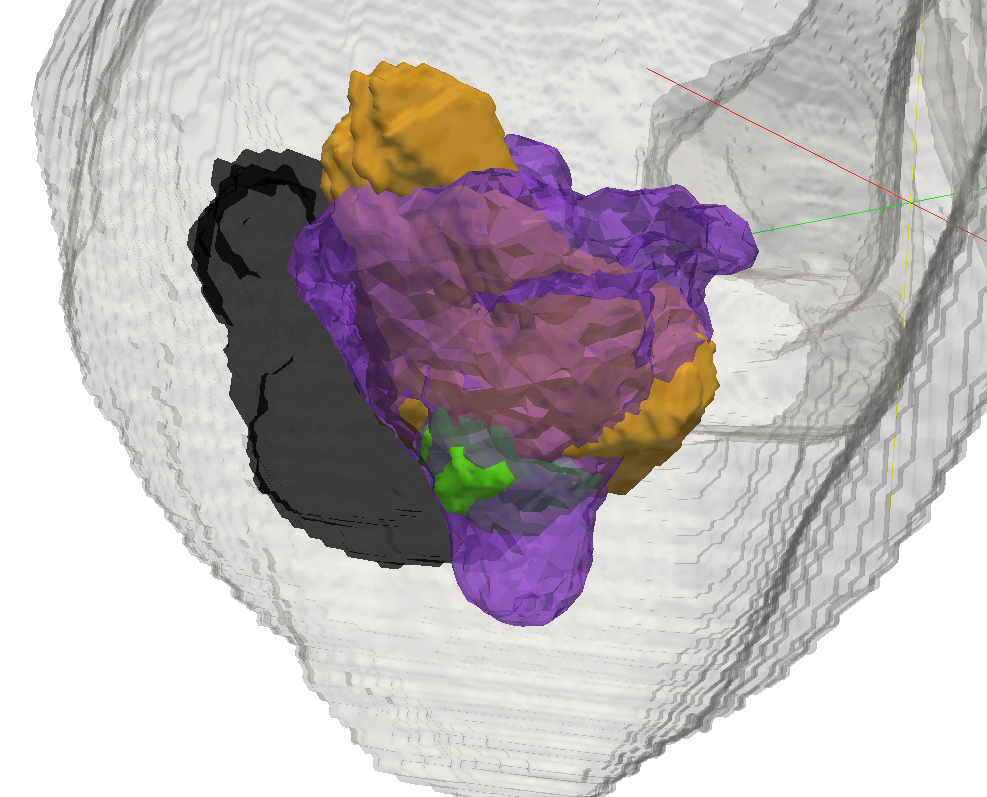}} &
    \resizebox{160px}{!}{\includegraphics{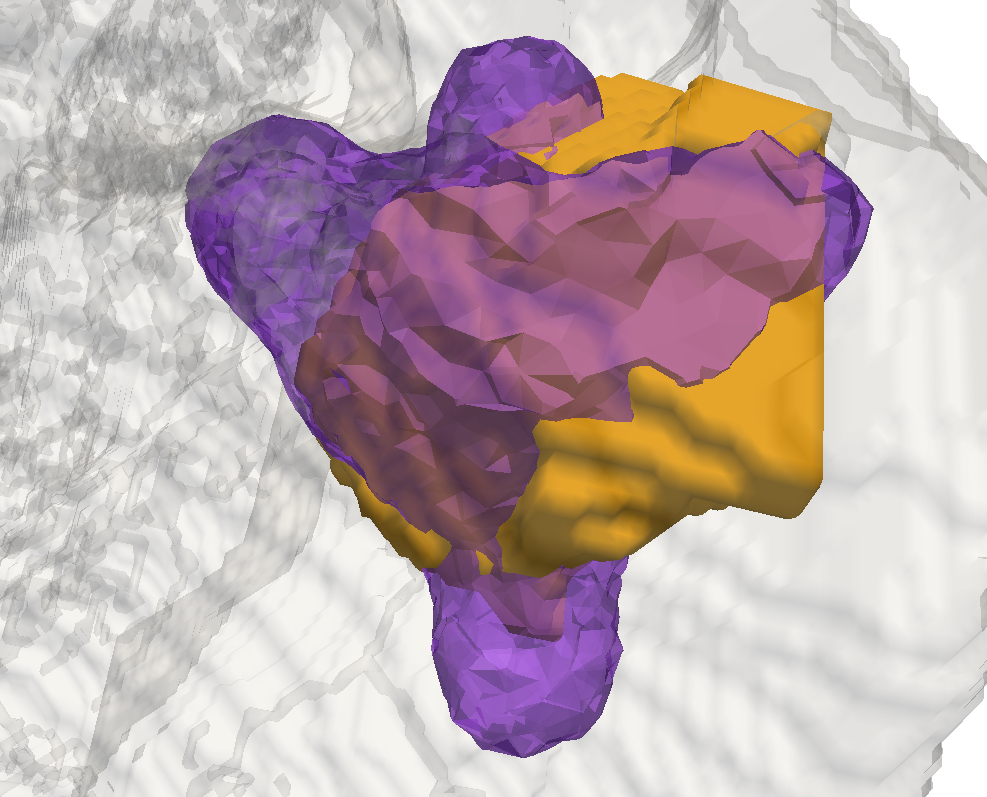}}\\
    (a) RFA1 & (b) RFA2
  \end{tabular}
  \caption{\label{fig-rfa-ablations}Radiofrequency ablation profiles for two
  cases. Purple: simulated lesion; orange: simulated lesion; green: tumour [in
  (a)]; charcoal: segmented lesion before rigid body registration [in (a)];
  light-grey outline: organ}
\end{figure}

The most and least severe deviations (as measured by AAE) are shown in Figure
\ref{fig-rfa-ablations}. In both cases, shape deviation can be seen between
the registered segmented ablation lesion in purple and the simulated ablation
lesion in organ. This similarity of shape deviation corresponds to the similar
observed AAE values. The simulated RFA heat deposition profile is based on
research outcomes of the IMPPACT project, and it is generally observed that
segmented RFA lesions form approximations to this pattern. Both lesions were
produced with Rita Starburst needles using a staged protocol, recommended by
the manufacturer, intended to produce a 5cm-diameter lesion. The clinical
steps and responses in this protocol are algorithmically defined in the
Go-Smart database.

Not shown in Figure \ref{fig-rfa-ablations}, to avoid obscuring the view of
the ablation lesions, is the vessel tree. The effects of this structure may be
seen to the far right of the simulated ablation in Figure
\ref{fig-rfa-ablations}(a), where the lesion profile curves inwards. The
percutaneous needle is not shown - in both cases it lies approximately on the
vertical axis, entering at the bottom of the figure and, in Figure
\ref{fig-rfa-ablations}(a), terminating inside the green tumour surface. The
individual flexible tines of the Rita Starburst probe extend at separate
stages of the protocol and lie a few millimetres inside the simulation extent
shown.

In Figure \ref{fig-rfa-ablations}(a), the clinically segmented lesion is shown
prior to rigid-body transformation (in charcoal). The transformed version is
shown in purple. The inaccuracy of the original 1-2 month follow-up
registration may be seen by the offset of the tumour, which is segmented in
the pre-interventional image, from the charcoal profile. In contrast, the base
of the simulated lesion mostly covers the tumour. However, there may be a
small local underestimation here, as no indication of under-treatment was
given in the clinical report form (CRF). For clarity, only the transformed
version is shown in most subsequent figures.

\subsection{Microwave ablation}

\begin{table}[h]
  \begin{tabular}{c|lllll}
    & Organ & DICE & SN & PPV & AAE ($\text{mm}$)\\
    \hline
    MWA1 & Liver & 0.503 & 0.349 & 0.903 & 3.682\\
    MWA2 & Lung & 0.545 & 0.466 & 0.657 & 5.305\\
    MWA3 & Liver & 0.722 & 0.579 & 0.958 & 2.285\\
    MWA4 & Liver & 0.722 & 0.603 & 0.901 & 2.278\\
    MWA5 & Lung & 0.650 & 0.517 & 0.870 & 4.071\\
    \hline
    $\mu$ &  & 0.628 & 0.583 & 0.859 & 3.524\\
    $\sigma$ &  & 0.090 & 0.091 & 0.104 & 1.15
  \end{tabular}
  \caption{\label{tab-eval-mwa}Evaluation measures for a series of MWA
  treatments}
\end{table}

All microwave ablation interventions presented in Table \ref{tab-eval-mwa}
were performed at the Medical University of Graz [AU] (Liver) and University
Hospital Frankfurt [DE] (Lung).

Cases MWA1-MWA5 generally have significantly larger PPV values than SV. All
three cases performed in the liver have PPV values over 0.9, although only two
have DICE values over 0.7. The liver cases have a smaller AAE, with values
between 2.2\,mm and 3.7\,mm, compared to the lung cases, with the AAE values
of both being over 4.0\,mm. This suggests that, both in the lung and in the
liver, the MWA model may be consistently over-estimating the extent of the
ablation lesion. The moderate AAE values in the liver suggest that there it is
more accurately reflecting the shape than in the lung. Overall, the mean,
$\mu$, is similar to the RFA cases, but the standard deviation, $\sigma$, is
much larger, approximately treble the presented RFA example.

\begin{figure}[h]
  \begin{tabular}{cc}
    \resizebox{160px}{!}{\includegraphics{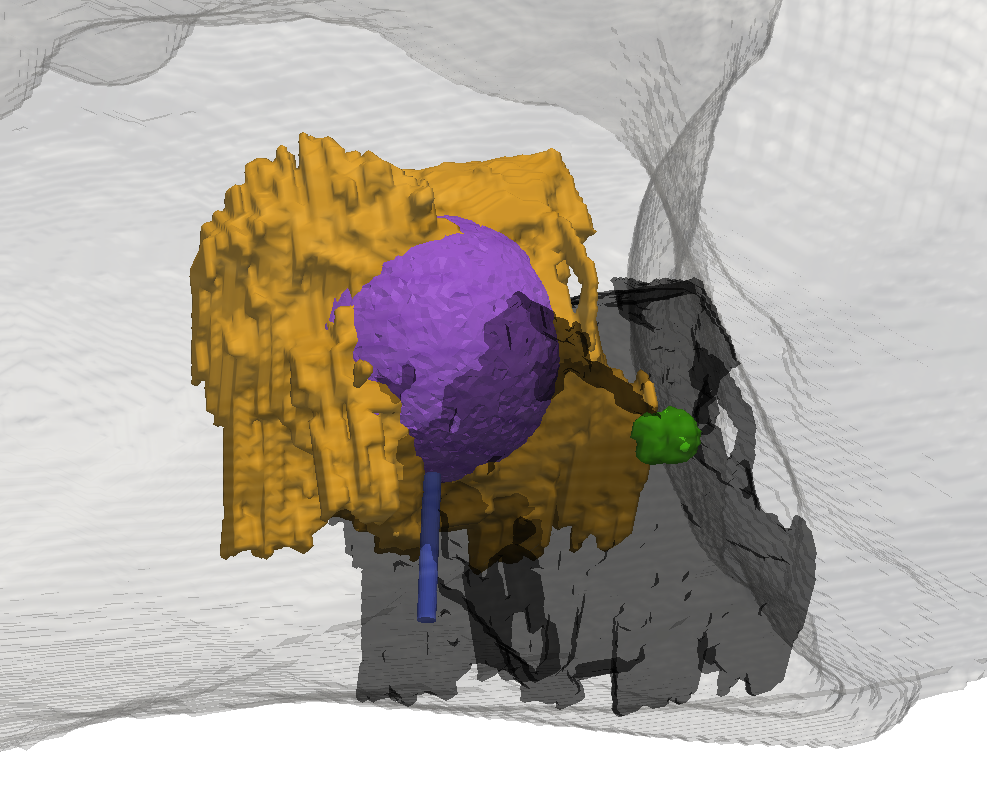}} &
    \resizebox{160px}{!}{\includegraphics{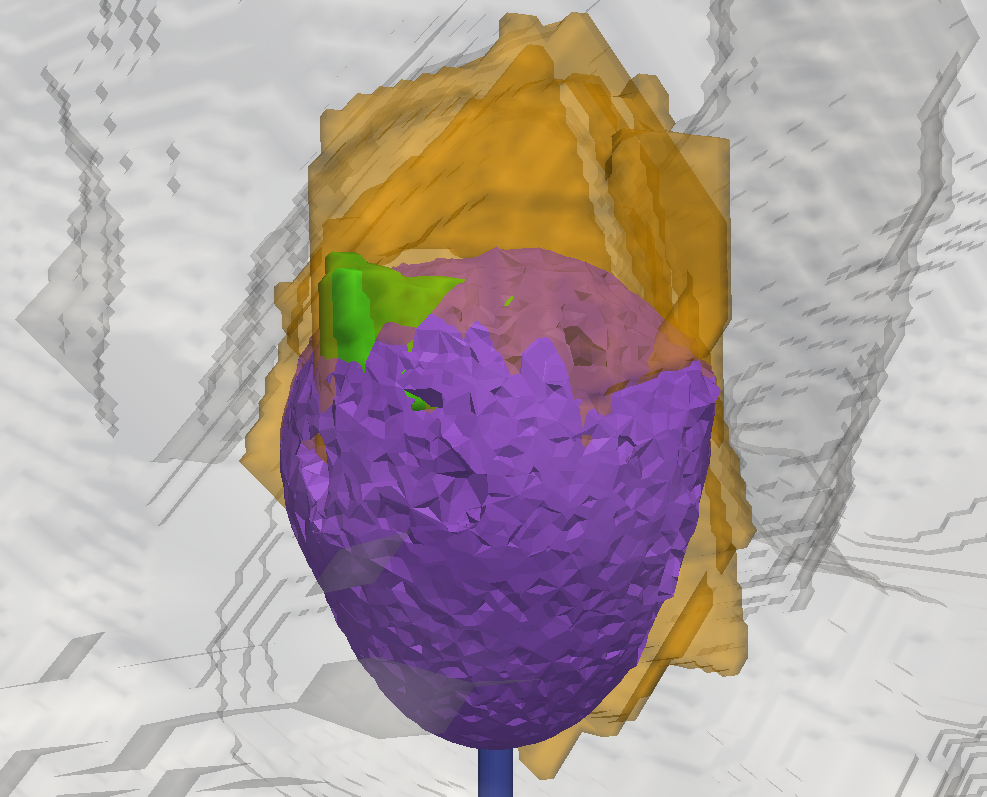}}\\
    (a) MWA2 & (b) MWA4
  \end{tabular}
  \caption{\label{fig-mwa-ablations}Microwave ablation profiles for two cases.
  Purple: simulated lesion; orange: simulated lesion; green: tumour; charcoal:
  segmented lesion before rigid body registration [in (a)]; light-grey
  outline: organ; blue cylinder: ablation needle}
\end{figure}

In Figure \ref{fig-mwa-ablations}, as in the RFA discussion, the two MWA cases
with greatest and least AAE values, respectively, are shown. Alongside the
vessel tree, an additional segmented structure is present in the lung
simulations, the bronchi (not shown). In both images, it can be seen that the
MWA profile forms an approximately obloid shape. While this is produced using
bioheat transfer and a deposition profile derived from Maxwell's
electromagnetic equations, the simulated ablation region is similar in profile
to descriptions in manufacturers' literature.

The case with greater deviation, MWA2, is shown in Figure
\ref{fig-mwa-ablations}(a). The clinician-segmented tumour and ablation zone
are significantly offset from the clinician-segmented needle. The simulation
itself matches the needle location well and in one dimension, has similar
extent to the segmented ablation lesion. However, in the directions
perpendicular to the line of view, the segmented ablation has a much larger
extent. In the second case, MWA4, shown in Figure \ref{fig-mwa-ablations}(b),
a much better match is visible. The tumour is mostly, but not entirely,
covered by the simulation. The ablation extent in line with the needle is
underestimated, however.

In the lung context, registration is considerably more challenging due to the
large deformations in the medium, and this may contribute to the discrepancy
observed in MWA2. In general, observed ablation profile deviations in the
liver tend to be smaller. In addition, the MWA energy deposition profile is
heavily dependent on the ablation probe's internal geometry, thus the
equipment model and manufacturer. While we have been able to produce models
based on prototypal probe geometries, this specificity is believed to
contribute to ablation underestimation in MWA4 and other liver and lung cases.
A second manufacturer-specific geometry is under development.

\subsection{Cryoablation}

\begin{table}[h]
  \begin{tabular}{l|lllll}
    & Organ & DICE & SN & PPV & AAE ($\text{mm}$)\\
    \hline
    CRYO1 & Kidney & 0.619 & 0.553 & 0.739 & 2.905\\
    CRYO2 & Kidney & 0.755 & 0.960 & 0.622 & 2.368\\
    CRYO3 & Kidney & 0.796 & 0.682 & 0.955 & 1.591\\
    CRYO4 & Kidney & 0.828 & 0.821 & 0.835 & 2.240\\
    CRYO5 & Kidney & 0.743 & 0.748 & 0.739 & 1.859\\
    \hline
    $\mu$ &  & 0.748 & 0.749 & 0.778 & 2.19\\
    $\sigma$ &  & 0.071 & 0.142 & 0.111 & 0.450
  \end{tabular}
  \caption{\label{tab-eval-cryo}Evaluation measures for a series of
  cryoablation treatments}
\end{table}

All cryoablation interventions presented in Table \ref{tab-eval-cryo} were
performed at the Radboud University Medical Centre [NE].

Performance in the CRYO1-CRYO5 cases is generally good, although no consistent
pattern of under- or over-estimation is easily observable. The AAE values are
generally low, although in CRYO1, all volumetric measures are below $0.74$.
CRYO5 has similar volumetric measures, but its AAE value is below 1.9,
suggesting that deviation may be a relatively small offset. In CRYO3, an even
lower AAE value is seen, but the SN value is the second lowest of the set,
perhaps indicating that the segmented ablation in this case is uniformly
larger than the simulated profile.

\begin{figure}[h]
  \begin{tabular}{cc}
    \resizebox{160px}{!}{\includegraphics{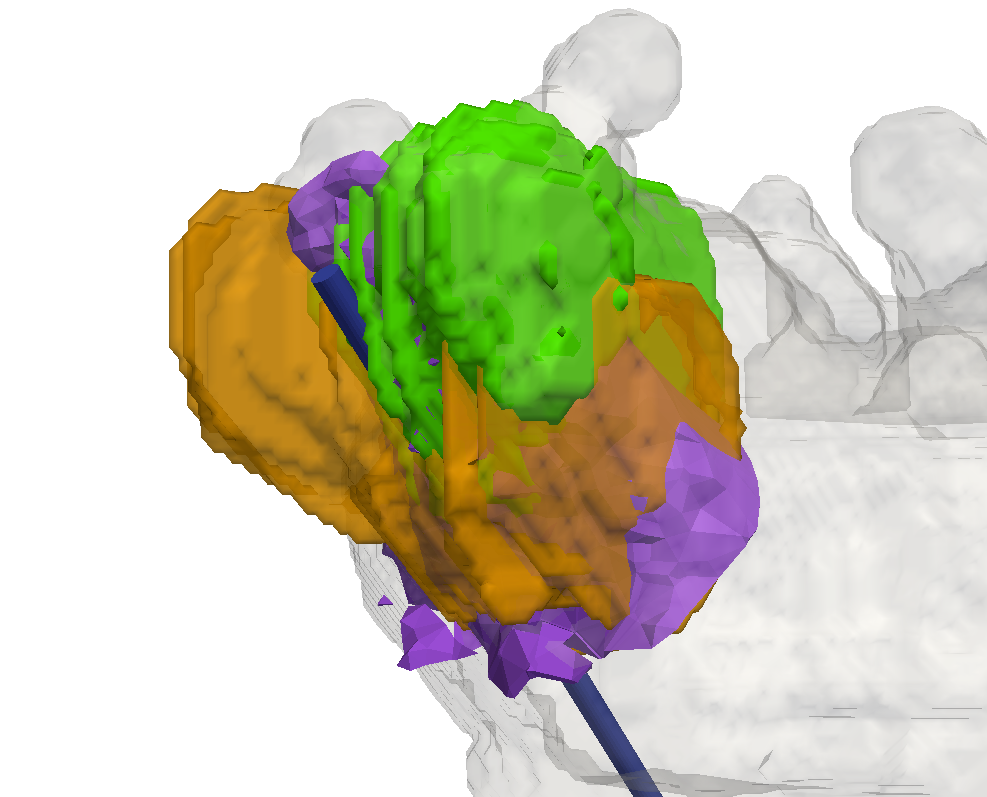}} &
    \resizebox{160px}{!}{\includegraphics{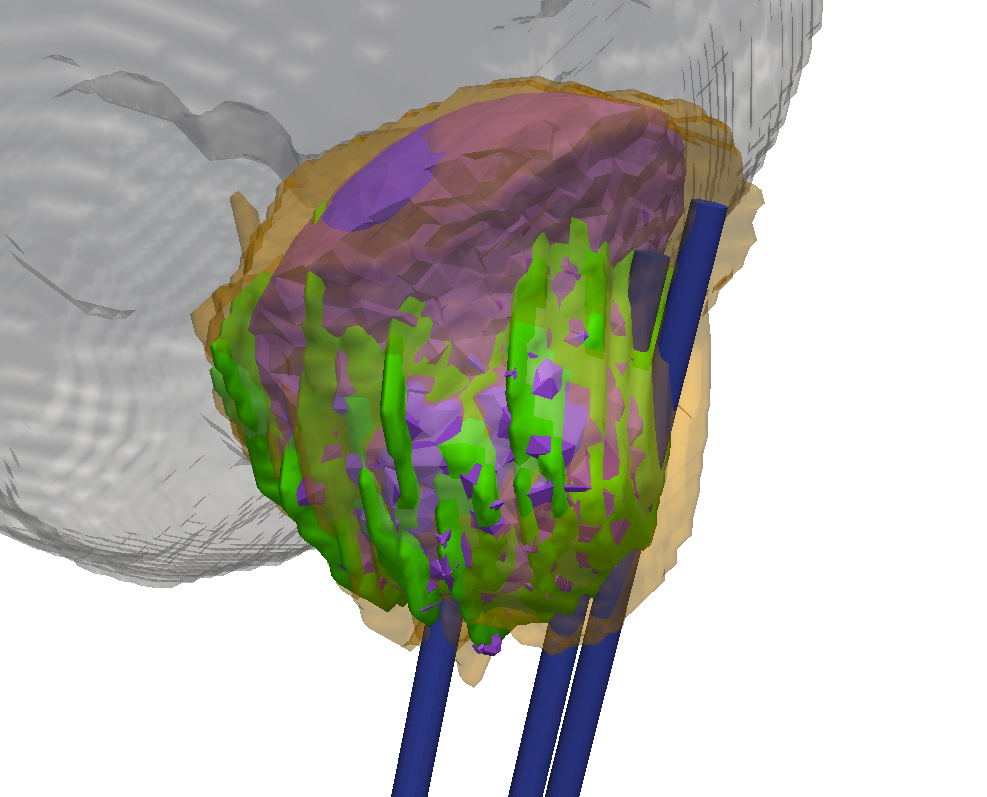}}\\
    (a) CRYO1 & (b) CRYO3
  \end{tabular}
  \caption{\label{fig-cryo-ablations}Cryoablation profiles for two cases.
  Purple: simulated lesion; orange: simulated lesion; green: tumour;
  light-grey outline: organ; blue cylinder: ablation needle}
\end{figure}

As before, in Figure \ref{fig-cryo-ablations}, the two cryoablation cases with
greatest and least AAE values, respectively, are shown. While MWA and RFA are
simulated using one needle embedded at a time, both cryoablation and IRE must
cater for multiple needles simultaneously embedded in the computational mesh,
as in Figure \ref{fig-cryo-ablations}(b).

While the sizes are broadly similar, deviation is visible in CRYO1 both in
shape and orientation. The tumour is not completely covered by either the
segmented or simulated ablation, although there is a region of three-way
overlap towards the base of the probe. Toward the tip, the probe itself leaves
the organ, as well as the lesion regions - while heat transfer is simulated
outside the organ in the cryoablation model, tissue necrosis is evaluated only
inside the segmented organ region. All cryoablation cases considered using
this tool are treatments of exophytic tumours and we see, both in CRYO1 and
CRYO3, that the organ wall provides a limit for the segmented lesion. CRYO3
also shows needles exiting the lesions and organ, although the match between
lesion extents is better.

The segmentation or registration may account for some of the difference
between CRYO1 and CRYO3 accuracy, as we can see the segmented lesion follows
the organ wall more accurately in the second case, but the simulation
nevertheless underpredicts the breadth of tumour. While the proximity to the
organ wall improves several aspects of the segmentation and registration
process, it introduces additional complication into the modelling process, as
the current model has little contextual information about the medium beyond
the organ wall. Future evolutions may include additional segmented regions
abutting the organ wall, allowing more accurate heat transfer modelling near
the needle tip.

\

\subsection{Irreversible electroporation}

\begin{table}[h]
  \begin{tabular}{c|lllll}
    & Organ & DICE & SN & PPV & AAE ($\text{mm}$)\\
    \hline
    IRE1 & Liver & 0.541 & 0.888 & 0.389 & 5.264\\
    IRE2 & Liver & 0.588 & 0.859 & 0.447 & 3.944\\
    IRE3 & Liver & 0.088 & 0.996 & 0.046 & 11.767\\
    IRE4 & Liver & 0.086 & 0.944 & 0.045 & 13.024\\
    IRE5 & Liver & 0.569 & 0.905 & 0.415 & 8.471\\
    \hline
    $\mu$ &  & 0.374 & 0.917 & 0.268 & 8.48\\
    $\sigma$ &  & 0.234 & 0.049 & 0.182 & 3.53
  \end{tabular}
  \caption{\label{tab-eval-ire}Evaluation measures for a
  series of irreversible electroporation treatments}
\end{table}

All irreversible electroporation interventions presented in Table
\ref{tab-eval-ire} were performed at the Medical University of Leipzig [DE].

The IRE1-IRE5 cases show a variety in quality. The PPV values in cases IRE3
and IRE4 are very low, but more acceptable in other cases. The SN values are
generally high with AAE inversely following the PPV, suggesting that the
segmented ablation zones generally lie inside the simulated zones. The
standard deviation in DICE and AAE, particularly, is much higher than in
previous cases, suggesting that cases may be mixed between adequate prediction
and inaccurate results.

\begin{figure}[h]
  \begin{tabular}{cc}
    \resizebox{160px}{!}{\includegraphics{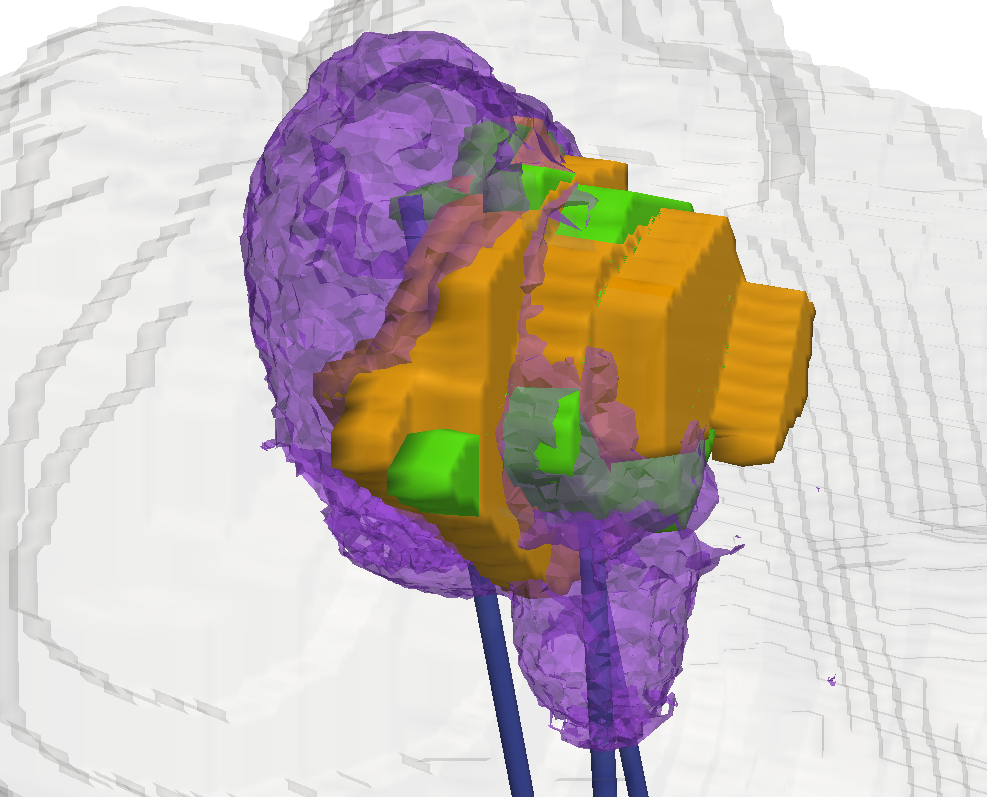}} &
    \resizebox{160px}{!}{\includegraphics{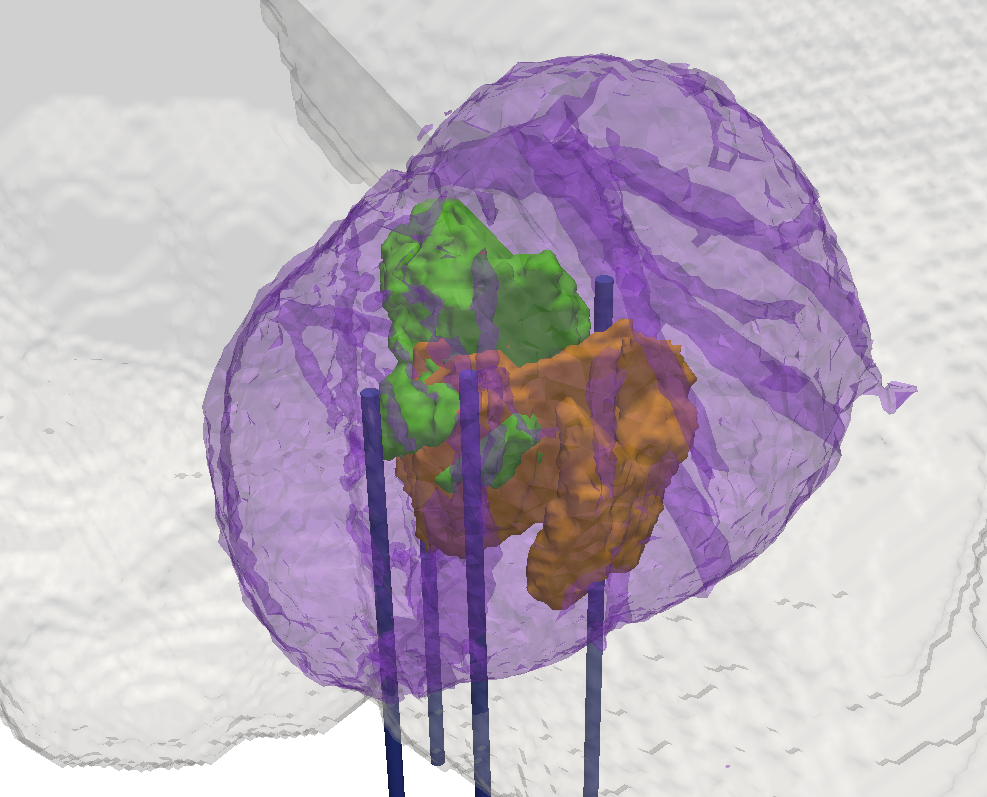}}\\
    (a) IRE2 & (b) IRE4
  \end{tabular}
  \caption{\label{fig-ire-ablations}Irreversible electroporation profiles for
  two cases. Purple: simulated lesion; orange: simulated lesion; green:
  tumour; light-grey outline: organ; blue cylinder: ablation needle}
\end{figure}

In Figure \ref{fig-ire-ablations}, case IRE2 and case IRE4 are shown. In IRE2,
we can see that for a small number of needles, close together, the
cross-sectional size is similar, although the extent lateral to the needles is
greater. In IRE4, the segmented ablation lesion lies mostly between the
needles, intersecting only two. It covers little of the tumour, although no
residual tumour tissue was identified in the follow-up acquisitions for this
case. While the simulated ablation does cover the tumour and needles, it is
considerably bigger than the segmented ablation.

In general, the IRE cases tend to match either acceptably well, or
significantly over-estimate the observed lesion. In the latter cases, a
segmented lesion disjoint from at least one needle is observed, an unlikely
occurence in the accepted physical theory, and consultation of the relevant
medical images confirms that the segmentation matches the measured data. A
possible explanation is that the necrotic tissue caused by the IRE ablation
beings the healing process faster than in thermal modalities. This is
supported by the absence of both tumour tissue and a visible ablation lesion
across much of the original tumour extent. Preliminary work has taken place to
compare simulated IRE lesions with lesion extents observed immediately
post-intervention, rather than 1-2 months later, showing much more favourable
comparison.

\subsection{Project outcomes}

The development of the Go-Smart tool gave rise to a number of key lessons that
were learned and incorporated:

\paragraph{Microservices architecture}To enable a scalable and resilient
service, deployable locally or in the cloud, components must be cleanly
articulated and re-usable.

\paragraph{Domain model spanning clinical and technical concerns}To facilitate
generic, open-ended simulation, relevant domain knowledge within the unified
model must not be from a single sphere, but reflect both practical and
theoretical modelling considerations.

\paragraph{Familiarity of interface}For a system to be usable by
interventional radiologists, the workflow must be a variation on standard
interfaces. In particular, simulation definition must fit neatly within
established tool layouts, and dynamically-added parameters should contain
definitions for user-friendly widgets and labelling.

\paragraph{Status updating}For longer, non-interactive steps, status updates
should be provided, both for technical users to analyse progress and
non-technical users to indicate continued processing. This necessitates a
live, multi-step communications relay from the containerized simulation,
through the simulation orchestrator, middleware and back to the client.

\paragraph{Broad clinical feedback is vital}Having interventional radiologists
using the system and analysing (pseudonymized) patient data within it is
essential to provide insight for technical users into clinical issues with the
system and to gather information on potential edge cases with patient imaging.

\paragraph{Interaction with manufacturers is needed to provide quality
models}Especially in multiphysics biomedical problems, manufacturer data and
feedback is required to ensure modelling starts from an accurate basis.

\paragraph{In trials, patient-specific modelling requires on-going engagement
between clinicians and modellers}While Clinical Report Forms are the
fundamental tool in contextualizing image data, there must be case-by-case
discussion between clinicians and technical users to identify shortcomings in
modelling, unpredicted issues and potential errors in parameter interpretation
or measurement. In particular, an expertise gap between numericists' awareness
of clinical received knowledge, and clinical understanding of biophysics
modelling concerns, may only be adequately addressed by in-depth discussion of
individual interventions.

\section{Conclusions}

This work has outlined a web-based platform, applied to image-guided Minimally
Invasive Cancer Treatments, that serves the needs of clinicians, manufacturers
and researchers simultaneously. To achieve this requires a conceptual model
spanning the clinical and modelling concerns. For researchers, this
multi-stage approach allows them to deploy new models or computational
routines to the web-based platform with minimal additional burden. For
manufacturers, this provides a tool to demonstrate their products and increase
awareness within the market. For clinicians, this provides a platform for
training, collaboration and knowledge gathering.

The models used for Minimally Invasive Cancer Treatment were shown to be
adequate for demonstration purposes and sufficiently varied to establish the
necessary flexibility of the Go-Smart platform. Improvements, both in terms of
speed and accuracy, may be made incrementally, and as a collaborative exercise
through the web interface. At present, our models produce most favourable
results for cryoablation, where bioheat transfer is the primary modelling
challenge, and RFA, where a heuristic approach has been established. In
microwave ablation, the sensitivity of the models to manufacturer-specific
geometry and complexity of electromagnetic modelling remain a challenge. In
IRE, the time-based definition of an ablation zone must be revisited.

Interaction between stakeholders of multiple groups is essential for
development of reliable and, ultimately, clinically useful models. A process
for facilitating this is required, and the refinements made during the
Go-Smart project in response to user feedback have helped streamline it as an
appropriate.

Producing a system that allows for extension of simulation capabilities by
end-users of the web-interface has been achieved at a lower level than we have
seen, although deployment of entirely new solver frameworks remains an
administrative task. More generally, the cross-concern conceptual model and
toolchain for containerized simulation orchestration is designed to be
re-usable in other clinical application domains.

\subsection{Future work}

On-going trials with external technical groups are underway, within a MICT
equipment manufacturer and a non-MICT-related biomedical research group.
Alongside clinical feedback, this will provide us with a useful corpus of
usability information to refine the current workflow.

To provide broader applicability, with low entry burden, existing open source
medical simulation tools, such as SimTK {\cite{sherman2005simtk}}, Chaste or
OpenCMISS, may be pre-emptively incorporated as new Glossia families. In
particular, wrapping Taverna or Galaxy as new families would allow much more
complex workflows to be built, and be executable within our current
architecture. Extending preliminary work (using
Vigilant\footnote{https://github.com/redbrain/vigilant}) to a more
user-friendly set of web-based back-end tools, providing orchestration
visualization and interactive log aggregation, will simplify management of
simulation deployments in a cloud setting.

\end{document}